\documentclass{aa}  
\usepackage{natbib,graphicx}
\usepackage{txfonts}
\usepackage{bm}
\usepackage{lmodern}
\usepackage{textcomp}
 \usepackage{array}  		
\usepackage{subfig}
\usepackage{amsmath}
\usepackage{amssymb}
\usepackage{hyperref}
\usepackage[final]{pdfpages}
\usepackage{media9}

\hypersetup{
  colorlinks=true,
  urlcolor=blue,
  citecolor= blue,
  linkcolor = blue
}

\begin{document}

   \title{Stellar streams as gravitational experiments}

   \subtitle{I. The case of Sagittarius}

   \author{Guillaume F. Thomas\inst{1}, B. Famaey\inst{1}, R. Ibata\inst{1}, F. L\"ughausen\inst{2} \and P. Kroupa\inst{2,3}}

   \institute{Universit\'e de Strasbourg, CNRS UMR 7550, Observatoire astronomique de Strasbourg, 11 rue de l’Universit\'e, F-67000 Strasbourg, France\\
              \email{guillaume.thomas@astro.unistra.fr}
              \and
              Helmholtz-Institut f\"ur Strahlen-und Kernphysik, Universit\"at Bonn, Nussallee 14-16, D-53115 Bonn, Germany
              \and
              Charles University in Prague, Faculty of Mathematics and Physics, Astronomical Institute, V  Hole\v{s}ovi\v{c}k\'ach 2, CZ-180 00 Praha 8, Czech Republic\\}

   \abstract{Tidal streams of disrupting dwarf galaxies orbiting around their host galaxy offer a unique way to constrain the shape of galactic gravitational potentials. Such streams can be used as ``leaning tower" gravitational experiments on galactic scales. The most well-motivated modification of gravity proposed as an alternative to dark matter on galactic scales is Milgromian dynamics (MOND), and we present here the first ever N-body simulations of the dynamical evolution of the disrupting Sagittarius dwarf galaxy in this framework. Using a realistic baryonic mass model for the Milky Way, we attempt to reproduce the present-day spatial and kinematic structure of the Sagittarius dwarf and its immense tidal stream that wraps around the Milky Way. With very little freedom on the original structure of the progenitor, constrained by the total luminosity of the Sagittarius structure and by the observed stellar mass-size relation for isolated dwarf galaxies, we find reasonable agreement between our simulations and observations of this system. The observed stellar velocities in the leading arm can be reproduced if we include a massive hot gas corona around the Milky Way that is flattened in the direction of the principal plane of its satellites. This is the first time that tidal dissolution in MOND has been tested rigorously at these mass and acceleration scales.}


\authorrunning{G. F. Thomas et al.}

   \maketitle

\section{Introduction}

The nature of the dark sector of the Universe represents one of the most pressing questions of modern physics. Over the years, we have built a large-scale picture in which the Universe is composed of only 5\% of ordinary baryonic matter and 25\% dark matter (DM), the rest being accounted for by a cosmological constant in the Einstein equation \citep{planckcollaboration_2016}. While arguably very succesful on large scales, this current $\Lambda$CDM picture is nevertheless plagued by a certain number of problems on small scales, especially on galaxy scales. Among those are the now famous `too-big-to-fail' and `satellite planes' problem \citep[e.g.,][]{kroupa_2005,kroupa_2010,boylan-kolchin_2011,papastergis_2015,pawlowski_2015}, but also a general fine-tuning problem encapsulated in the relation between the surface density of baryons and the gravitational field in galaxies. This relation is now often referred to as the {\it Radial Acceleration Relation} \citep[RAR][]{mcgaugh_2016,lelli_2016}, and involves an acceleration constant $a_0 \simeq 10^{-10} {\rm m}\,{\rm s}^{-2}$: this relation is also refelected in the tight baryonic Tully-Fisher relation \citep{mcgaugh_2000, lelli_2016a, papastergis_2016}, in the relation  between  the  stellar  and  dynamical  surface  densities  in  the central regions of galaxies \citep{lelli_2016b,milgrom_2016a}, in the relation between the central rotation curve slope and the baryonic surface density \citep{lelli_2013,renaud_2016}, or in the diversity of shapes of rotation curves at a given maximum velocity scale \citep{oman_2015}. While there have been attempts to explain the RAR in the classical DM picture \citep[e.g.,][]{navarro_2016}, these are far from convincing for a variety of reasons. For instance, \citet{navarro_2016} significantly overpredict the mass discrepancies, and assume, to get there, a tight stellar mass-size relation, whereas the latter is not actually observed. In fact, as shown by the comprehensive study of \citet{desmond_2017}, when using realistic correlations in the galaxy-halo connection, the RAR scatter is still overpredicted even when the abundance matching scatter is switched off \citep[see also][for comparisons of $\Lambda$CDM galaxy simulations with the RAR]{wu_2015}. One possible explanation for this conundrum would be that gravity is effectively different in the extremely weak field regime, and accounts for the effects usually attributed to particle DM in galaxies. This hypothesis is known as Modified Newtonian Dynamics (MOND), or Milgromian dynamics \citep{milgrom_1983}, which has predicted all  the aforementioned galaxy scaling relations (and pushed observers to look for them), and especially the RAR, well before they were precisely assessed by observations \citep{famaey_2012}. However, such a description has, of course, also its own problems. Some faint dwarf spheroidals deviate from the predicted relation \citep{mcgaugh_2010} which would mean that they are out of equilibrium in this context, or that the paradigm must be extended. Some tensions also exist in globular clusters which behave in a seemingly Newtonian way when MOND would a priori predict a deviation from it \citep{ibata_2011}. Even more problematic is the need for dissipationless non-baryonic DM to reproduce the angular power spectrum of the cosmic microwave background \citep{planckcollaboration_2016}, as well as the failure of the MOND relation in galaxy clusters, needing either dissipationless DM or additional baryonic DM. Various hybrid frameworks have thus been proposed \citep[e.g.,][]{blanchet_2015a,berezhiani_2015} in which a new degree of freedom can in principle play the role of cosmological DM on large scales, while effectively mediating precisely a MOND force in galaxies. However, those have not yet been shown to be as succesful as $\Lambda$CDM on large scales, and they are of course less minimal. Other theories posit that a scalar field is responsible for dark energy, and it has been proposed that this scalar field may interact with dark matter. The scalar field may mediate additional long-range forces between dark matter particles of comparable strength to the canonical gravitational force, although the theory currently makes no predictions about the strength of the additional force. Of course, since the dark matter particles accelerate differently to baryons, this implies that the weak equivalence principle would have to be broken. \citet{kesden_2006a,kesden_2006} argued that disrupting satellite galaxies could allow one to investigate this effect.

Here we concentrate on MOND, a modification of gravity whose predictions are known to be quite succesful in galaxies, without dark matter \citep{famaey_2012}. However, these predictions have been mostly limited until recently to rather symmetric and static configurations \citep[but see, e.g.,][for a few exceptions]{brada_1999,tiret_2007,tiret_2008,angus_2014,nipoti_2007}. This has changed with the advent of numerical codes, in particular the recent patches to the RAMSES code \citep{teyssier_2002} developed by \citet{lughausen_2014} and \citet{candlish_2015}. The {\it Phantom of Ramses} patch \citep{lughausen_2014} has for instance recently led, following the seminal work of \cite{tiret_2008a} on the topic, to the first MOND simulations of galaxy encounters with a detailed Eulerian hydrodynamical treatment of gas physics, including star formation and stellar feedback, finding that the star formation history is significantly more extended in time and space in MOND encounters than in the classical case \citep{renaud_2016a}. 

Other very powerful dynamical probes of the gravitational potential, which have not yet been thoroughly investigated in MOND, are the tidal stellar streams of disrupting satellite galaxies. These are especially important probes of the three-dimensional shape of the potential outside of the Galactic plane, and at large distances. Such an analysis is not straightforward because streams do not delineate orbits and because the non-linear external field effect of MOND is likely to play a role, hence the availability of a MOND N-body code is crucial for correctly tackling the problem. This is precisely what we aim to achieve here with this series of papers on stellar streams in modified gravity.

The stream of the Sagittarius (Sgr) dwarf galaxy \citep{ibata_1994} is the most prominent stellar structure around the Milky Way (MW) and the one for which we have data of the most exquisite precision. The orbit of the Sgr dwarf around the Galaxy is nearly polar, and the resulting tidal stream wraps a full 360$^\circ$ on the sky \citep{ibata_2001}. The detailed investigations of the stellar stream and its kinematics have led to a lot of confusion about the corresponding shape of the gravitational potential \citep{helmi_2004,johnston_2005,law_2005}. No model to date is satisfactory, and the benchmark model to compare with is still the one of Law \& Majewski (2010, hereafter LM10) based on the spatial and kinematic structure of M-giant stars of the stream \citep{majewski_2003,majewski_2004}, which relies on a triaxial almost oblate ellipsoid for the DM halo, but with its minor axis contained within the Galactic plane, which is not natural in $\Lambda$CDM \citep{debattista_2013,pearson_2015}.  In the context of a scalar field mediating a long-range force on dark matter particles, Kesden \& Kamionkowski also undertook a series of simulations of the Sagittarius dwarf galaxy in which they studied the influence of changing the strength of gravity for the dark matter. If the additional force is stronger than Newtonian gravity, the dark matter particles accelerate faster into the MW potential leaving the stars slightly behind. Because of this, any stars that do leave the system during the tidal disruption process are more likely to leave through the L2 Lagrange point, so that the resulting star stream appears asymmetric, with a less populated leading arm than the trailing arm. With observations of the stellar stream of the Sagittarius dwarf available at that time, \citet{kesden_2006a,kesden_2006} were able to rule out a 9\% higher acceleration for the dark matter.

More than a decade ago, \citet{read_2005} had shown that the orbit of the Sgr dwarf in a Milky Way MOND potential was barely distinguishable from that in a nearly-spherical to mildly oblate DM halo. However, as the stream does not follow the orbit, and as non-linearities in the MOND Poisson equation (notably the `external field effect' breaking the Strong Equivalence Principle, see Section 6.3 of Famaey \& McGaugh 2012) can lead to {\it a priori} unexpected effects, and it is thus urgent to revise this problem by using our modern simulation tools. Let us emphasize that a framework such as MOND has very little freedom to achieve the right final configuration of the stream, hence this exercise has a huge potential for falsification of the paradigm. Indeed, one must start from a progenitor dwarf sitting on the observed mass-size relation, and then hope that the global shape of the stream, the structure and kinematics of the remnant, and the total luminosity of the stream all fit with observations: all this with a gravitational potential of the MW fully determined by its baryonic distribution. This is not a trivial task.

We describe our method in Section~2 hereafter, then run in Section~3 two Newtonian simulations for comparison purposes, and finally run two MOND simulations of the Sgr stream in Section~4. We conclude in Section~5.

\section{Method}

In the following study, all N-body simulations were made with the RAMSES code \citep{teyssier_2002}. For simulations in the MOND framework, we used the {\it Phantom-Of-Ramses} (POR) patch developed by \citet{lughausen_2014}, who generalized the Poisson equation in the following way \citep{milgrom_2010}:
\begin{equation}
\label{eqn:1}
\nabla^2\Phi =  \nabla.\left[\nu\left(\frac{|\nabla \Phi_\mathrm{N} |}{a_0}\right) \nabla \Phi_\mathrm{N} \right] \, ,
\end{equation}
where $\Phi$ and $\Phi_\mathrm{N}$ are the MOND and Newtonian potentials respectively, $\nu(x) =1$ for $x \gg 1$ (Newtonian regime) and $\nu(x)=x^{-1/2}$ for $x \ll 1$ (deep-MOND regime),  and $a_0 = 1.2 \times 10^{-10}$ m.s$^{-2}$ is the acceleration constant of the MOND paradigm.
We can then introduce the ``phantom dark matter'' density $\rho_\mathrm{ph}$ such that
\begin{equation}
\label{eqn:poisson}
\nabla^2\Phi = 4\pi G \left[\rho_\mathrm{b} + \rho_\mathrm{ph} \right],\end{equation}
which is fully defined through Eq.~\ref{eqn:1} once the baryonic distribution $\rho_\mathrm{b}$ (and its associated Newtonian potential $\Phi_\mathrm{N}$) is known, and can be seen as the MOND equivalent to the DM contribution in the classical case. This phantom DM density, that does not correspond to real particles, is computed at each time-step in POR and used to compute the MOND potential. The code is based on Adaptative-Mesh-Refinement (AMR) that increases the resolution of the grid in higher density regions without a dramatic growth in computation time. In our specific case the resolution will be highest along the tidal stream and in its progenitor. Since the energy along an orbit is conserved in a static external potential, it is important to keep the same resolution in the inner region of the satellite during the whole simulation time to avoid a numerical modification of the energy of the progenitor that could modify the orbit. In this work, we chose a minimum resolution of the AMR grid of 31~kpc and a maximum resolution of 15~pc.

In the following sections, we will proceed as follows: first we will devise benchmark Newtonian models for comparison with our MOND results, based on a spherical halo and the triaxial halo of LM10. These will be labelled simulaitions N1 and N2. These simulations are presented purely for comparisons with the following MOND simulations, and should thus not be over-interpreted in the CDM context. We will then move on to MOND simulations, with and without a massive hot gaseous corona around the Galaxy (simulations M1 and M2, respectively). Let us note here that the behavior of streams in MOND depends both on their own internal gravitational field and on the external field of the host galaxy. For these reasons, there is no general `simple' test case to present here, as each situation will actually be different based on the internal properties of the progenitor and the properties of the host and of the orbit.

All our models are based on the following baryonic matter distribution for the MW \citep{dehnen_1998}: a double exponential stellar disk of $3.52 \times 10^{10} M_\odot$ for the thin and thick disk components, with a scale length of 2 kpc and two scales heights of 0.3 and 1~kpc. The bulge and the interstellar medium components have respectively a mass of $0.518 \times 10^{10}$ M$_\odot$ and $1.69 \times 10^{10} M_\odot$. This MW model is not live in the following simulations, and is represented by $5.6 \times 10^{5}$ static particles of 10$^5$ M$_\odot$ each, that generate the static potential in RAMSES. Hereafter, we will call this distribution of matter the \textit{disk model}.
 
In all our simulations, we will follow the disruption of the Sgr dSph for 4 Gyr. The stream is indeed composed of relatively `young' M-giants and is dynamically young \citep{majewski_2003}. Moreover, we will assume that the MW does not have an important modification of mass due to major mergers or heavy accretions, and that its morphology stayed the same during the last 4 Gyr. To determine the orbit of the progenitor and the initial position of the dSph in Newtonian dynamics, it is common to launch a point mass test-particle from its present location and make it run backwards, assuming a progenitor with a negligible mass compared to the mass of the host galaxy, i.e. neglecting dynamical friction with the DM halo. This is reasonable if the progenitor is less than $\sim 10^9 M_\odot$, which is the case for the Sgr dSph galaxy during the last 4 Gyr in the case of a spherical DM halo \citep{penarrubia_2006}, but which is in stark contradiction with abundance matching, which requires the mass of the dwarf galaxy to be at least $10^{10.5} M_\odot$. This tension could actually be problematic for CDM-based models  \citep[\citeauthor{kroupa_2015} \citeyear{kroupa_2015} but see][]{dierickx_2017}. But if one ignores this tension and stays with a smaller mass, the dynamical friction is also negligible in the case of a DM halo with a similar triaxiality as that of LM10, as we will show hereafter. In MOND, this problem is of course trivially circumvented, as there is no dynamical friction outside of the MW disk  \citep[but note however, as a caveat, that within the disk, dynamical friction is actually more efficient in MOND than in Newtonian dynamics, which would thus be very important for, e.g., in-plane accretions][]{ciotti_2004, nipoti_2008}. 

The present-day position and velocity of the Sgr dSph are listed in Table \ref{param_Sgr}, where the distance and the radial velocity are heliocentric. We use the same proper motion as LM10, however we choose to adopt here the Solar peculiar motion of \citet{schonrich_2010}, namely $(U_\odot, V_\odot, W_\odot) = (11.1, 12.24, 7.25)\, {\rm km}\,{\rm s}^{-1}$ in Local Standard of Rest coordinates, and a Sun-Galactic Center (GC) distance of 8.5~kpc. Also the present-day apparent magnitude of the Sgr dSph is $m_V = 3.63$ \citep{mateo_1998,majewski_2003} corresponding to an absolute magnitude of  $M_V = -13.64$, i.e. a V-band total luminosity of $2.4 \times 10^7$ L$_\odot$. The half-light radius along the minor axis is $r_h= 0.6$ kpc and the central velocity dispersion is $\sigma_c=11.4$ ${\rm km} \, {\rm s}^{-1}$ \citep{majewski_2003,ibata_1997}, as summarized in Table~\ref{param_Sgr}. These quantities here refer to the remnant dSph only and do not include the associated stream. The initial conditions for our four simulations N1, N2, M1 and M2, as obtained by integrating a test-particle backwards in time for 4 Gyr, are given in Table~\ref{param_model}.

  \begin{table}
 \centering
  \caption{Properties of the Sgr dSph remnant in terms of position, velocity, half-light radius along the minor axis $r_h$, central velocity dispersion, and total luminosity. The sources are : $1 =$ \citet{ibata_1994}, $2 =$ \citet{law_2010}, $3 =$ \citet{ibata_1997}, $4 =$ \citet{majewski_2003}. The value of the distance and luminosity of the galaxy differs from the value of LM10 since we chose a Sun-GC distance of 8.5 kpc.}
  \label{param_Sgr}
  \begin{tabular}{@{}ccc@{}}
  \hline 
   Parameter & Value & Source  \\
    \hline 
   RA & $18^h 55^m 19.5^s$ & 1 \\
   Dec & $-30\degr 32'43.0"$ & 1 \\
   Distance & $28.5$ kpc & 2 \\
   $\mu_\alpha$ & -2.45 mas.yr$^{-1}$ & 2 \\
   $\mu_\delta$ & -1.30 mas.yr$^{-1}$ & 2 \\
   V$_{rad}$ & $+140 \pm 2.0$ ${\rm km} \, {\rm s}^{-1}$ & 3\\
   $r_h$ & 0.6 kpc & 3\\
   $\sigma_c$ & 11.4 ${\rm km}\,{\rm s}^{-1}$ & 3\\
   $L_V$ & $2.4 \times 10^7$ L$_\odot$ & 4\\
\hline
\end{tabular}
\end{table}  

\begin{table*}
 \centering
  \caption{Initial positions and velocities of the Sgr dSph in our different simulations, to match the position and velocities of the remnant (as listed in Table~\ref{param_Sgr} after 4 Gyr, $X=19.0$, $Y=2.7$, $Z=-6.9$ kpc, $Vx=231.6$, $Vy=-40.3$ and $Vz=200.0$ km.s$^{-1}$, where the coordinates $(X, Y, Z)$ are Galactocentric and in the right handed coordinate system.}
  \label{param_model}
  \begin{tabular}{@{}ccccccc@{}}
  \hline
    Model & X (kpc) & Y (kpc) & Z (kpc) & Vx (${\rm km} \, {\rm s}^{-1}$) & Vy (${\rm km} \, {\rm s}^{-1}$) & Vz (${\rm km} \, {\rm s}^{-1}$)   \\    \hline \\
    N1 & -53.93 & -11.54 & 45.37 & -25.74 & 25.30 & -67.74 \\
    N2 & -37.05 & -24.18 & 51.34 & -84.06 & 27.60 & -64.41 \\
    M1 & -30.71 & 1.00 & -4.19 & 205.01 & 34.63 & -138.96 \\
    M2 & -16.60 & 11.41 & -35.40 & 146.88 & 43.51 & -139.14 \\  
    \hline
\end{tabular}
\end{table*}

\section{Newtonian simulations with dark matter}

\begin{figure*}
\centering
  \includegraphics[angle=0, viewport= 10 15 1140 570, clip, width=17cm]{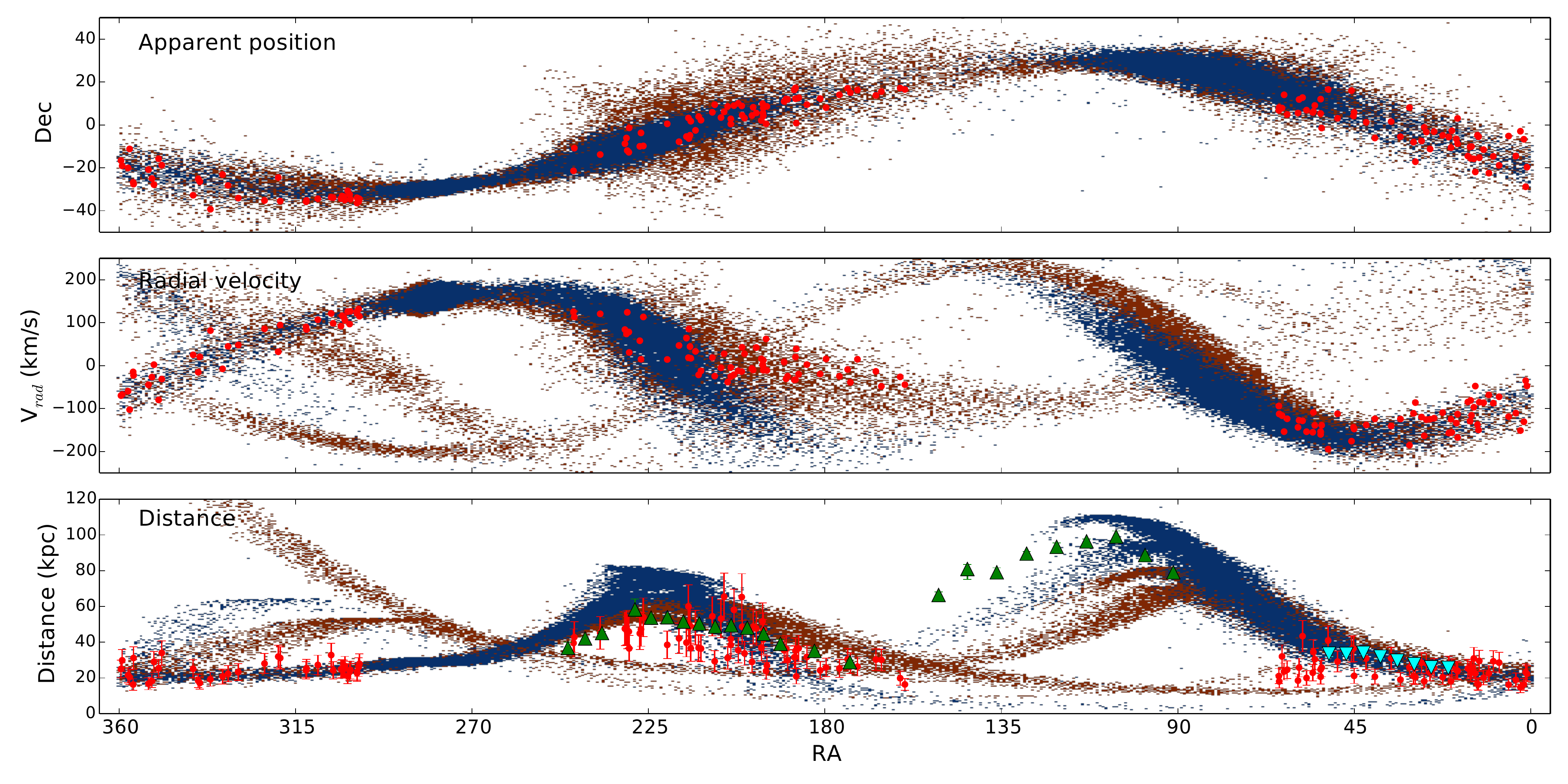}
  \caption{Projection of the N-body particles of the Sgr stream after 4 Gyr of disruption of a light progenitor in Newtonian dynamics, in blue for model N1 (spherical DM halo) and in orange for N2 (a triaxial halo as in LM10).  The red dots are the observed 2MASS M-giants stars from \citet{majewski_2004}, the green triangles are the BHB stars from \citet{belokurov_2014} and the cyan triangles correspond to the bright stream of \citet{koposov_2012}. The apparent position in equatorial coordinates are shown on the top panel, the heliocentric radial velocities in the middle panel and the heliocentric distances in the bottom panel. }
\label{proj_spherique_dm}
\end{figure*}

In this section, we first run two simulations in Newtonian dynamics for testing the code and for comparison purposes, first in the case of a spherical DM halo (simulation N1), and then in the case of the LM10 triaxial DM halo (simulation N2).

\begin{figure}
\centering
  \includegraphics[angle=0, viewport= 0 0 571 545, clip, width=7.5cm]{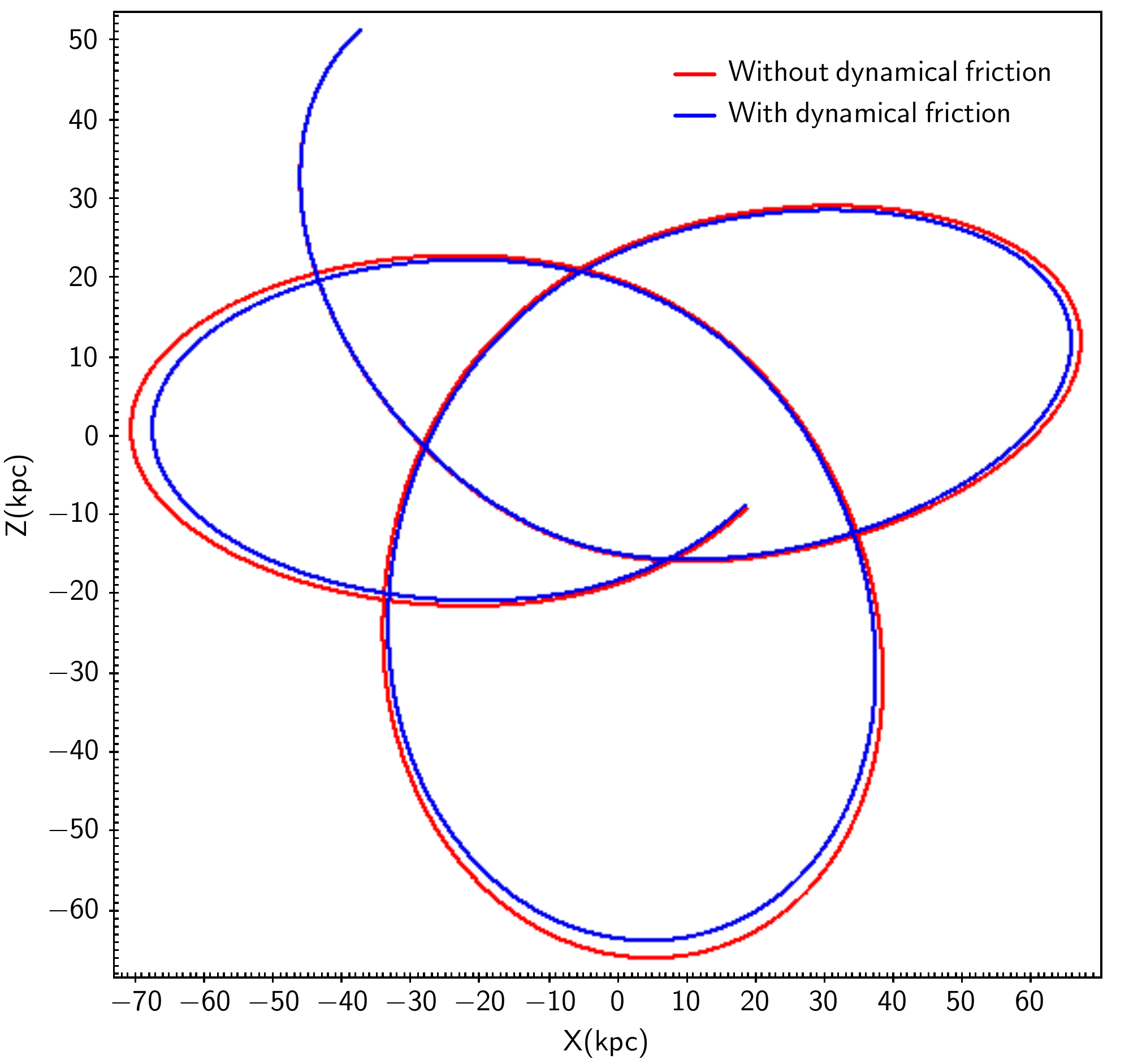}
  \caption{Orbit of the light Sgr dSph ($6.8 \times 10^{8} $ M$_\odot$) in Newtonian dynamics with the LM10 halo without dynamical friction (simulation N2) in red and with dynamical friction estimated through Chandrasekhar's formula in blue with Coulomb logarithm $\rm{ln}(\Lambda)=3$. The difference between the two orbits is clearly not very significant for such a light progenitor.}
\label{orbit_fricdyn}
\end{figure}

In simulation N1, we add to the MW disk model a spherical DM halo based on the double-power-law model of \citet{dehnen_1998} with $\rho_{h0} = 2.46 \times 10^8$ M$_\odot$.kpc$^{-3}$, $\alpha = -0.87$, $\beta=2.36$, a$_h = 2.66$ kpc and r$_h = 1000$ kpc that corresponds to M$(r<100 {\rm kpc}) = 6.7 \times 10^{11}$  M$_\odot$ and a virial mass of M$_{vir} = 1.5 \times 10^{12}$ M$_\odot$. This halo is modelled as an external potential with $1.5 \times 10^{6}$ static DM particles of individual mass 10$^6$ M$_\odot$. For the progenitor dwarf, the initial positions and velocities are given in Table~\ref{param_model}, and the internal structure is given by a light (DM+stars) King profile \citep{king_1966,binney_2008} with $M = 4 \times 10^{8}$ M$_\odot$, a core radius of $r_c = 0.65$~kpc and a ratio between the central velocity dispersion and the potential of $W = 5$. After running the simulation for 4~Gyr, the resulting apparent positions on the sky, heliocentric radial velocities and heliocentric distances of the particles are shown on Fig.~\ref{proj_spherique_dm} together with the observed 2MASS M-giant stars of \cite{majewski_2004}, taking their estimate of $\sim$ 20\% error on the distance. In this simple N1 case, while the apparent positions and distances are reasonably well-reproduced, this is not the case of the radial velocities in the bright leading arm of the stream between RA$= 140^\circ$ and $200^\circ$. The modelled velocities typically reach $-210 \, {\rm km}\,{\rm s}^{-1}$ while the observations do not go below $-50 \, {\rm km}\,{\rm s}^{-1}$. This means that the modelled particles fall back {\it too fast} towards the Galactic plane in this leading arm. This is actually a known problem since the work of \citet{law_2005}, who showed that the problem remains for an oblate or a prolate DM halo. These observed radial velocities can for instance also be compared to the recent model of \citet{dierickx_2017}: as can be seen on their figure~10, the problem is at least as severe in terms of radial velocities, whilst apparent positions on the sky are worse. One possibility is to discard these data as non-members of the stream. Another is to consider an alternative halo, which is precisely what LM10 did with their proposed triaxial halo. Here we reproduce their results with simulation N2 in Fig.~\ref{proj_spherique_dm}. Following e.g., \citet{penarrubia_2010}, we modelled this halo with a NFW profile with virial mass M$_{vir} = 1.1 \times 10^{12}$ M$_\odot$, substituting the spherical radius with an elliptical radius $m$ where $m^2 = x^2/a^2 + y^2/b^2 + z^2/c^2$ and $a=0.44,\,b=1.0, \,c=1.0$, similar to LM10. The progenitor dwarf here is modelled with a King profile with a mass $M_{init} = 6.8 \times 10^{8} $ M$_\odot$, a core radius $r_c = 0.65$~kpc and a ratio between the central velocity dispersion and the potential $W = 4$. Taking into account the loss of mass over 4~Gyr, we also use Chandrasekhar's formula \citep{chandrasekhar_1943} to estimate the effect of dynamical friction on this model: the difference is shown in Fig.~\ref{orbit_fricdyn} where it is clear that for a {\it light} Sgr dwarf, the dynamical friction does not modify the orbit significantly. With such a triaxial halo, the kinematics of stars in the bright leading arm at $140^\circ<{\rm RA}<200^\circ$ is reproduced, but note that the projected dispersion on the sky is clearly overproduced and that the stream is less extended in distance (Fig.~\ref{proj_spherique_dm}, bottom panel). Also, \citet{debattista_2013} have shown that such a halo is not able to host a stable disk, and recently, \citet{pearson_2015} have shown that due to this triaxiality, the stars of the Palomar 5 \citep[Pal 5, see e.g.][]{thomas_2016} stream would be on chaotic orbits and create a fanning shape at the end of the stream, which is not observed: they show that a spherical DM halo is better at reproducing the thin and coherent observed structure of the Pal~5 stream. We will return to the topic of modelling Pal~5 in modified gravity in a subsequent contribution. 

\section{MOND simulations}

Contrary to the above Newtonian dynamics case, in which there is significant freedom to choose the potential (because it is dominated at the distance of the Sgr dSph by the DM halo), the baryonic mass distribution of the host galaxy has a crucial importance in MOND, as it fully fixes the gravitational potential. The same is actually true for the progenitor dwarf galaxy: it must sit on the observed stellar mass-size relation, and at the end of the simulation, the remnant should resemble what is observed: again, there is very little freedom here, as the gravitational potential of the progenitor is entirely determined by its baryonic content. Finally, there is an additional effect, which is unique to MOND, which we would like to investigate here: the `external field effect'. This effect is absolutely unique to theories like MOND which  break the strong equivalence principle, and are distinct from the usual tidal effects. It means that the internal dynamics of a satellite system does not decouple from the external field produced by its host system, drastically reducing the amount of `phantom dark matter' at pericenter compared to the apocenter or isolated case, and even producing pockets of negative phantom DM densities at places. It is this effect that led to the successful prediction of the small velocity dispersion of the dwarf galaxy Crater II in MOND \citep{mcgaugh_2016,caldwell_2016}. One of the things we would like to understand is whether this effect plays a role in shaping stellar streams in MOND, and whether this provides a distinct signature from Newtonian gravity.

For our study, we need to choose a transition function $\nu$ in Eq.~1, between the MONDian and the Newtonian regime around the $a_0$ acceleration scale. The shape of this transition will not be absolutely crucial in our study, since the MW potential is already in the deep-MOND regime at the distance of the Sgr dSph. Following \citet{famaey_2005} and \citet{zhao_2006}, for galaxies\footnote{See however \citet{hees_2016} for tight constraints in the Solar System for the strong gravitation regime}, we choose an interpolation function of the form:
\begin{equation}
\nu(x) = \frac{1+(1+4x^{-1})^{1/2}}{2} \, .
\label{interpolation_simple}
\end{equation}

\begin{figure*}
\centering
  \includegraphics[angle=0, viewport= 0 50 560 520, clip, width=8.0cm]{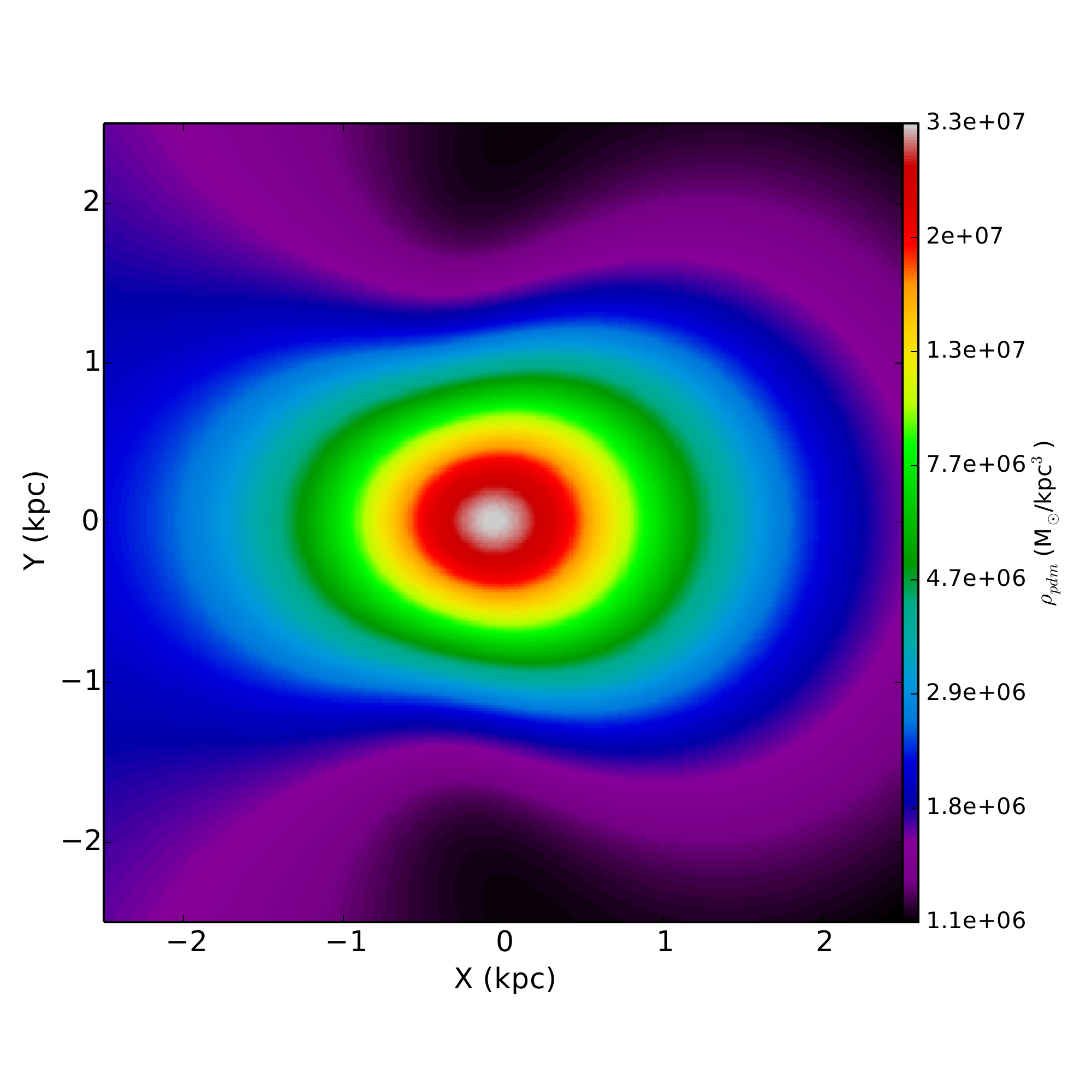}
  \includegraphics[angle=0, viewport= 0 50 560 520, clip, width=8.0cm]{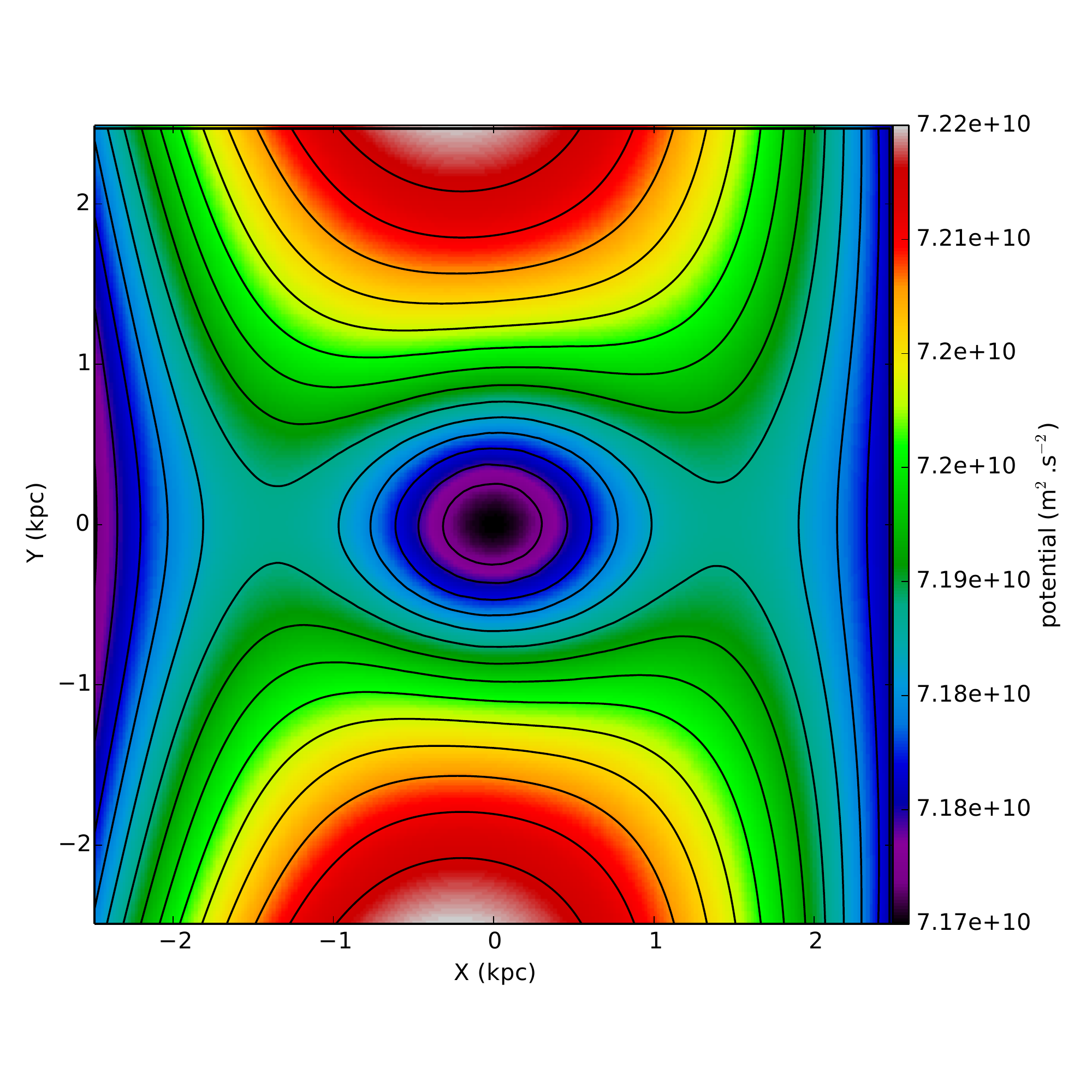}
  \includegraphics[angle=0, viewport= 0 50 560 520, clip, width=8.0cm]{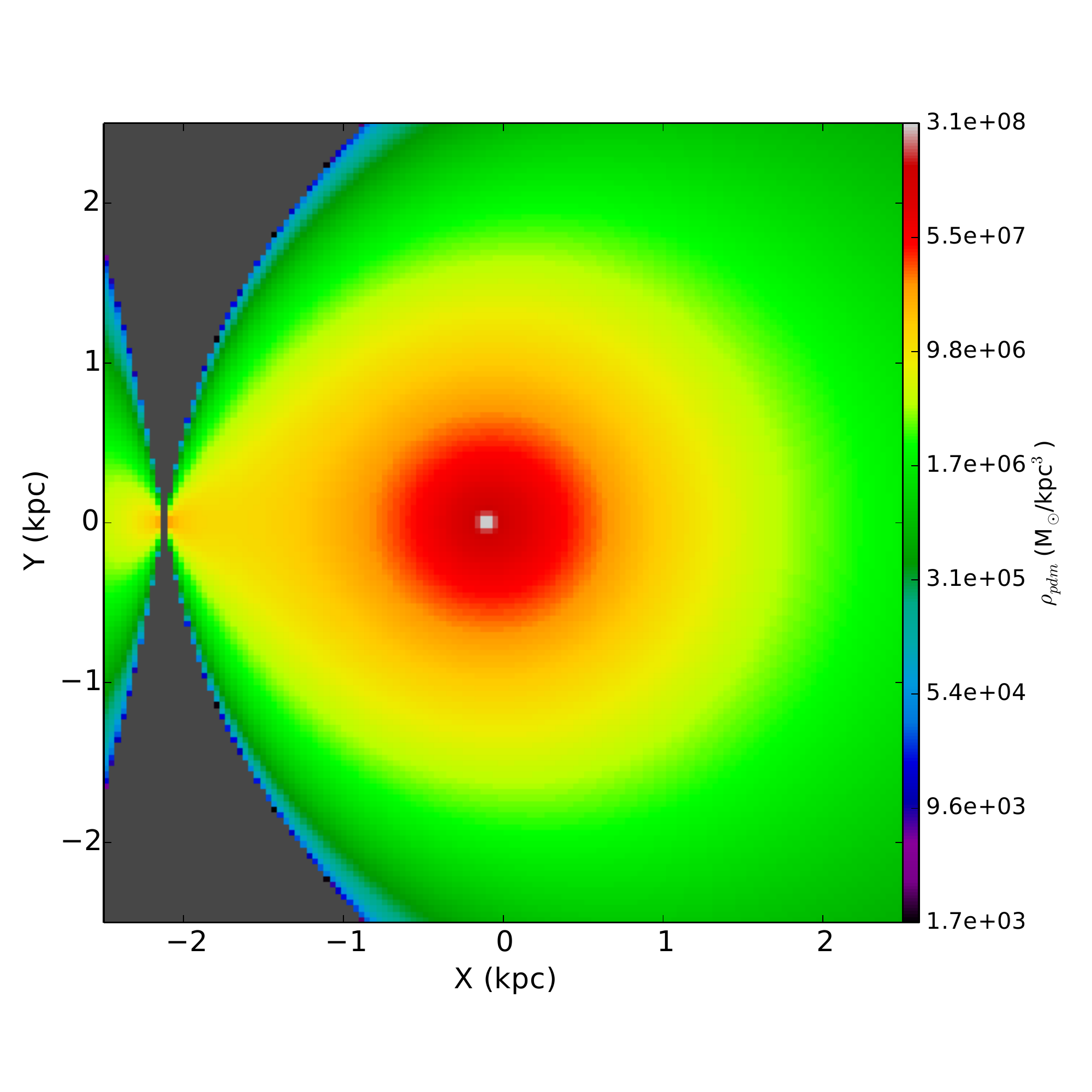}
  \includegraphics[angle=0, viewport= 0 50 560 520, clip, width=8.0cm]{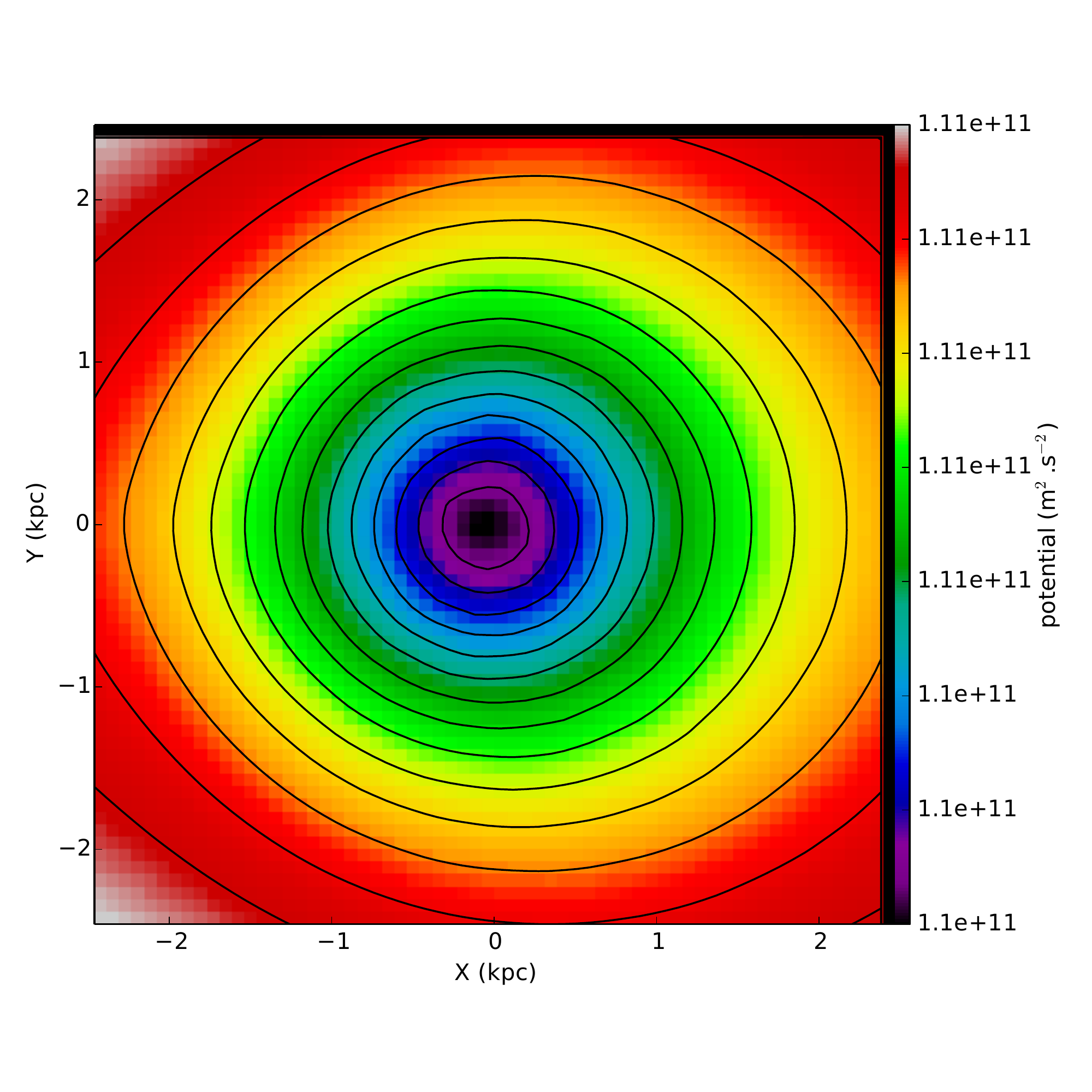}
  \caption{The two left panels display a cut of the density of phantom dark matter around a Plummer sphere with a Plummer radius $= 0.85$kpc ,  mass$=5.1 \times 10^7$ M$_\odot$) representing a Sgr-like progenitor at 20 kpc from the Galactic center (upper left panel) and at 80 kpc (lower left panel). The Galactic center is respectively at (X,Y) = (-20,0) kpc and (-80,0) kpc on this plot. The two right panels display the corresponding effective potential for a dwarf galaxy on a circular orbit around a $5.6 \times 10^{10}$ M$_\odot$ point mass.}
\label{pdm}
\end{figure*}

To estimate the external field effect, we first computed analytically the density of phantom dark matter around a Plummer sphere with Plummer radius of $r_s = 0.85$ and mass of $5.1 \times 10^7$ M$_\odot$, corresponding roughly to the current baryonic mass of the Sgr remnant assuming a mass-to-light ratio of $\sim 2.1$, on a circular orbit around a point mass of $5.6 \times 10^{10}$ M$_\odot$ at a distance of 20~kpc and 80~kpc, hence roughy corresponding to the expected pericenter and apocenter of the Sgr dwarf. As shown in Fig.~\ref{pdm} the distribution of phantom-dark-matter around the Plummer sphere is not spherical due to the external field effect \citep{wu_2010}. There are even pockets of negative density which compress the dwarf. At apocenter, the phantom-dark-matter mass is slightly larger, and so is its negative phantom density counterpart, but this happens beyond $\sim 2$~kpc from the center of the dwarf, and thus it does not affect the stream. The disruption of the satellite appears to be influenced mostly by the shape of the potential of the progenitor at pericenter, which is not lopsided and is very similar to the Newtonian case with dark matter, as can be seen on the top-right panel of Fig.~\ref{pdm}.

After these analytical preliminaries, we will now model the progenitor stellar distribution with a King profile, which is closer to the shape of the observed stellar distribution of dSph galaxies, and produce full N-body simulations of the Sgr disruption in MOND.

\subsection{Modellling the Sgr dSph in MOND} \label{model_sgr}

For the initial conditions on the internal kinematics of the dwarf, we constructed a fully self-consistent MONDian King model of a dwarf galaxy sitting on the observed stellar mass-size relation \citep{dabringhausen_2013}. It will be represented in our simulation by $10^5$ N-body particles. The King model \citep[see Section 4.3 of][for details]{binney_2008} is defined by a distribution function depending on energy, which once integrated over velocity space, gives a density $\rho_k$ proportional to the relative binding potential $\psi$:
\begin{equation}
 \rho_{k}  \propto e^{\psi / \sigma}  erf\left( \frac{\sqrt{\psi}}{\sigma} \right) - \sqrt{\frac{4 \psi}{\pi \sigma^2}} \left( 1 + \frac{2 \psi}{3 \sigma^2} \right) \, .
 \label{king_dens}
\end{equation}
To construct a MOND King model, all we need to do is replace the Newtonian relative potential $\psi_N$ of the classical formulation  by the MONDian relative potential $\psi$ in that equation, and we integrate from inside out, assuming that the central region is in the Newtonian regime. We halt the inside-out integration once outside of the desired radial range, and check with POR that the King model is indeed in equilibrium in isolation in MOND.
  \begin{table*}
 \centering
  \caption{Initial (t=0) and final (t=4~Gyr) stellar mass, half-light radius and central velocity dispersion of the Sgr dwarf in the two MOND simulations M1 and M2.}
  \label{param_progenitor}
  \begin{tabular}{@{}c|ccc|ccc@{}}
  \hline
    & \multicolumn{3}{|c|}{Initial progenitor} & \multicolumn{3}{|c}{Final remnant} \\
     
    Model & Mass (M$_\odot$) & $r_h$ (kpc)& $\sigma_c$ (km.s$^{-1}$)  & Mass (M$_\odot$) & $r_h$ (kpc) & $\sigma_c$ (km.s$^{-1}$)  \\
    \hline
    & & & & \\
    M1 & $1.2 \times 10^{8}$ & 0.61 & 24.0 & $5.1 \times 10^{7}$ & 0.64 & 11  \\
    M2 & $1.4 \times 10^{8}$ & 0.62 & 25.0 & $5.7 \times 10^{7}$ & 0.66 & 11\\
    \hline
\end{tabular}
\end{table*}

\subsection{The Sgr stream in the disk model}

In this section, we present a benchmark simulation of the Sgr stream in MOND, denoted M1. In this simulation, the only source of gravity is the (non-live) MW model presented in Sect.~2 and the live fully baryonic self-consistent King model devised with the method outlined above. 

Our first constraint will be to reproduce the total luminosity of the Sgr structure (the Sgr stream + the Sgr dSph) , which is of the order of (or a bit less than) $L \sim 10^{8}$ L$_\odot$ \citep{niederste-ostholt_2010}. After a few tries, we choose an initial stellar mass of $1.2 \times 10^{8} M_\odot$ for the model M1. We then choose a core radius such that the half-light radius matches the observed stellar mass-size relation of other dwarf galaxies \citep{dabringhausen_2013}. For this, we choose a core radius of $r_c = 0.6$ kpc and $W = 5$ for the ratio between the central velocity dispersion and potential. This leads to a central velocity dispersion of $\sigma_c = 24$ km.s$^{-1}$ and a half-light radius of $r_h = 0.61$ kpc (see Table~\ref{param_progenitor}). As we know that dynamical friction of the dwarf galaxy with an inexistent DM halo does not take place in MOND, we can safely integrate the orbit backwards in time in the MW disk model MOND potential to get the initial positions and velocities listed in Table~\ref{param_model}. We then run our M1 N-body simulation forwards in time for 4~Gyr.

\begin{figure}
\centering
  \includegraphics[angle=0, viewport= 6 9 531 396, clip, width=8.0cm]{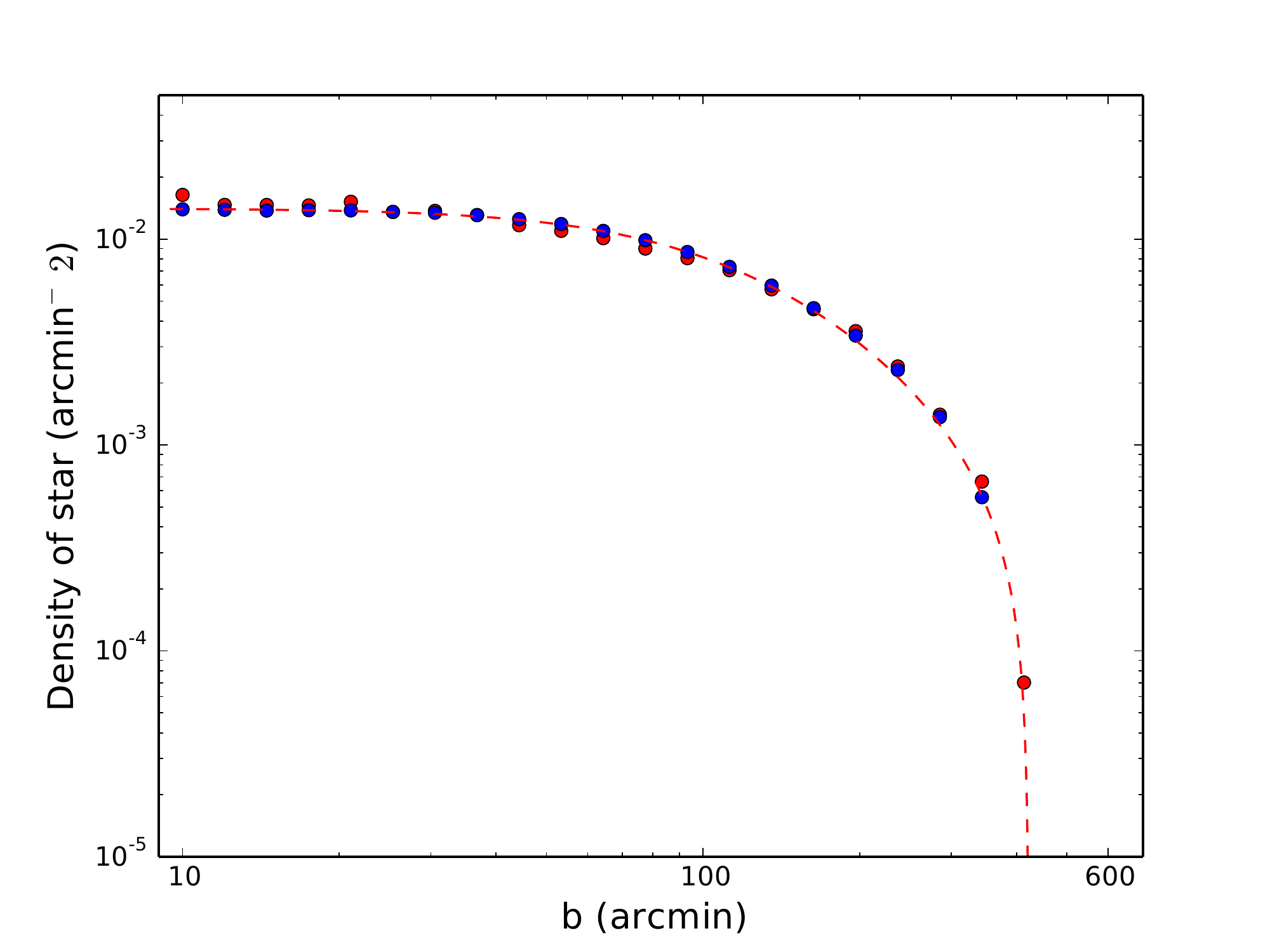}
  \caption{Stellar density along the minor axis of the Sgr dSph, of observed M-giant stars from the 2MASS survey (in red), and of the particles from the MOND disk model M1 simulation after 4 Gyr of integration (in blue). The red dashed line represents the best fit of a King profile to the observed M-giants.}
\label{profil_remnant}
\end{figure}

\begin{figure}
\centering
  \includegraphics[angle=0, viewport= 45 80 500 280, clip, width=8.0cm]{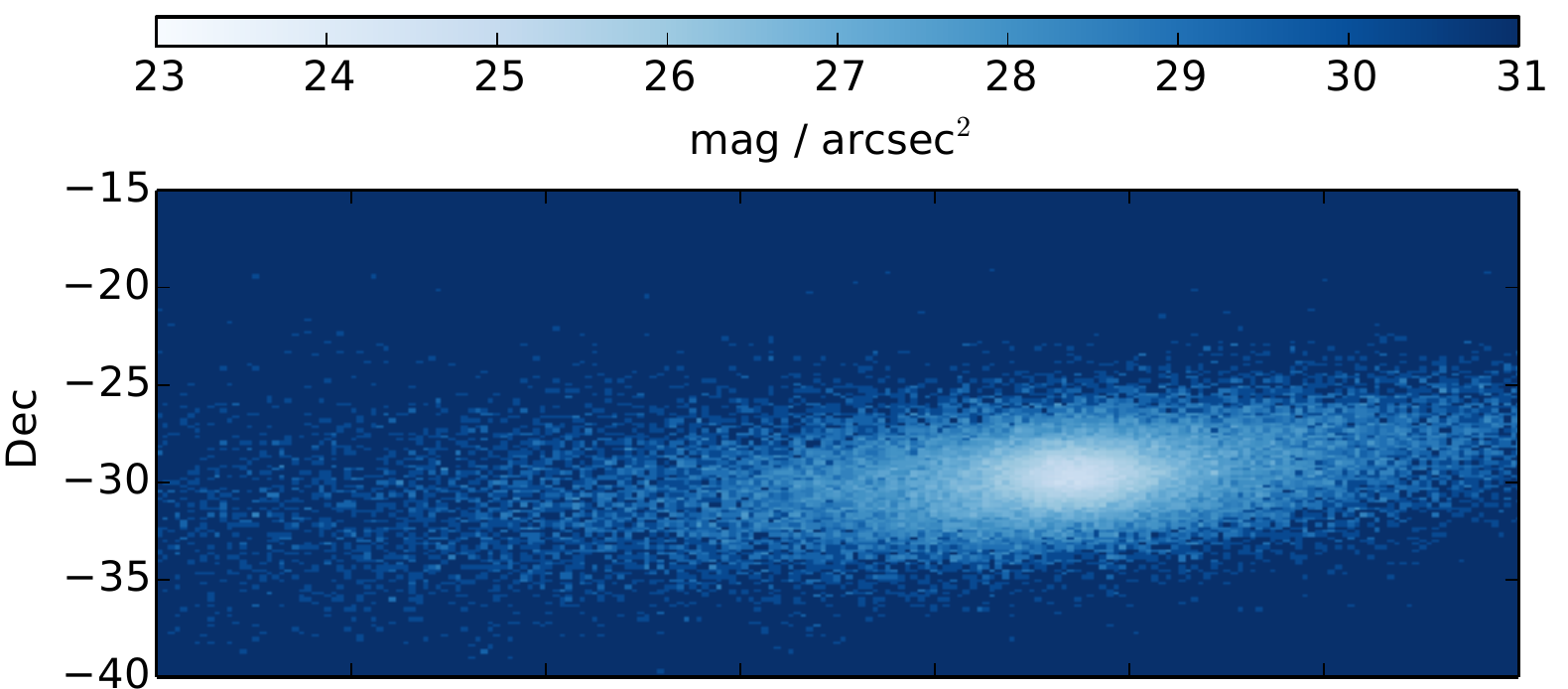}
  \includegraphics[angle=0, viewport= 45 53 500 234, clip, width=8.0cm]{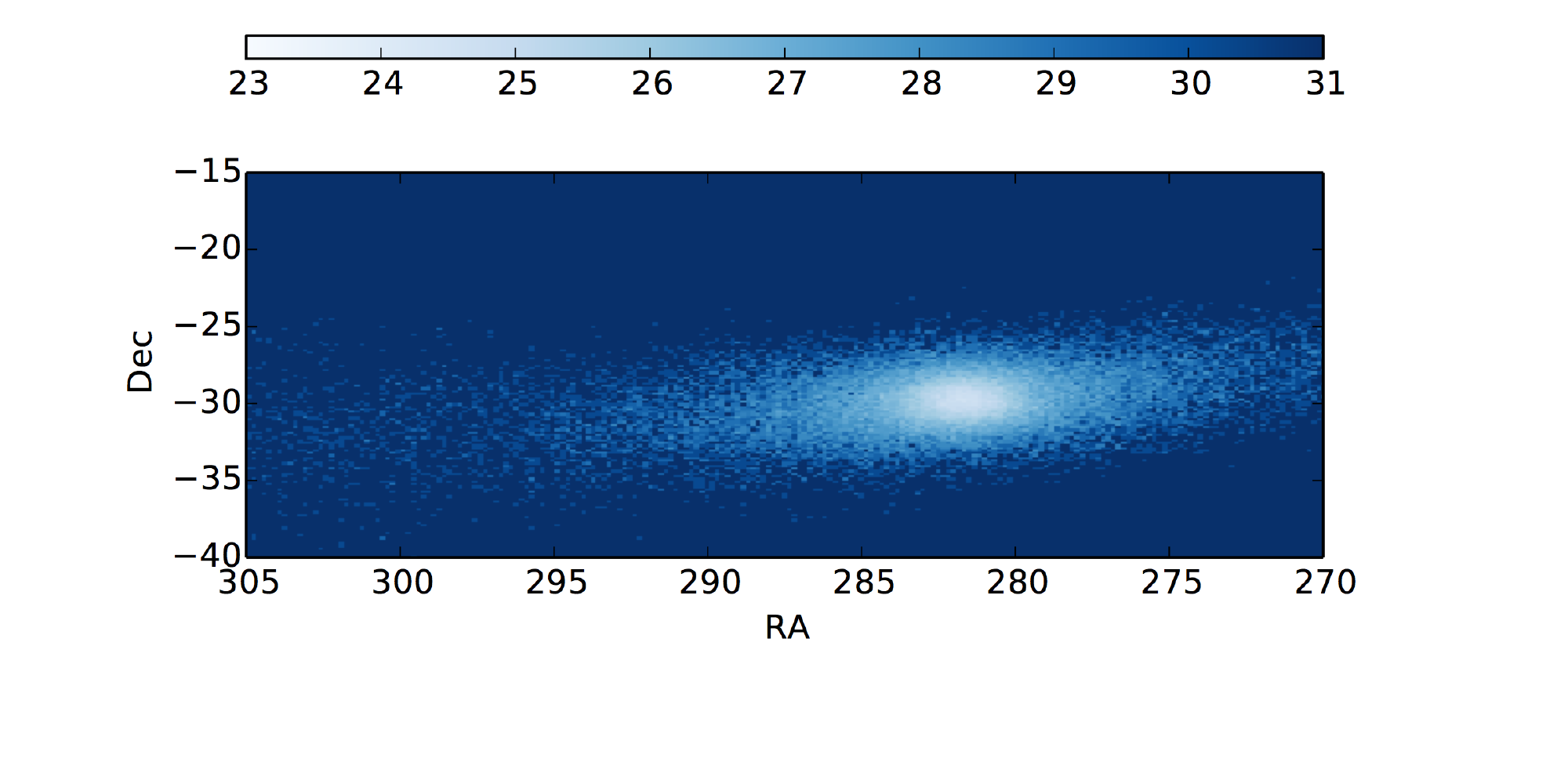}
  \caption{Surface brightness of the remnant in our simulations, assuming a stellar mass-to-light ratio in the \textit{V}-band of $\gamma_* = 2.1 $  for the M1 model on the upper panel and  $\gamma_* = 2.4 $ for the M2 model in the lower panel. The surface brightness in both cases is $\mu_0 \approx 24.6$ mag/arcsec$^2$. Compare this plot to Figure~4 of \citet{majewski_2003}.
}
\label{surface}
\end{figure}

\begin{figure}
\centering
  \includegraphics[ width=8.0cm]{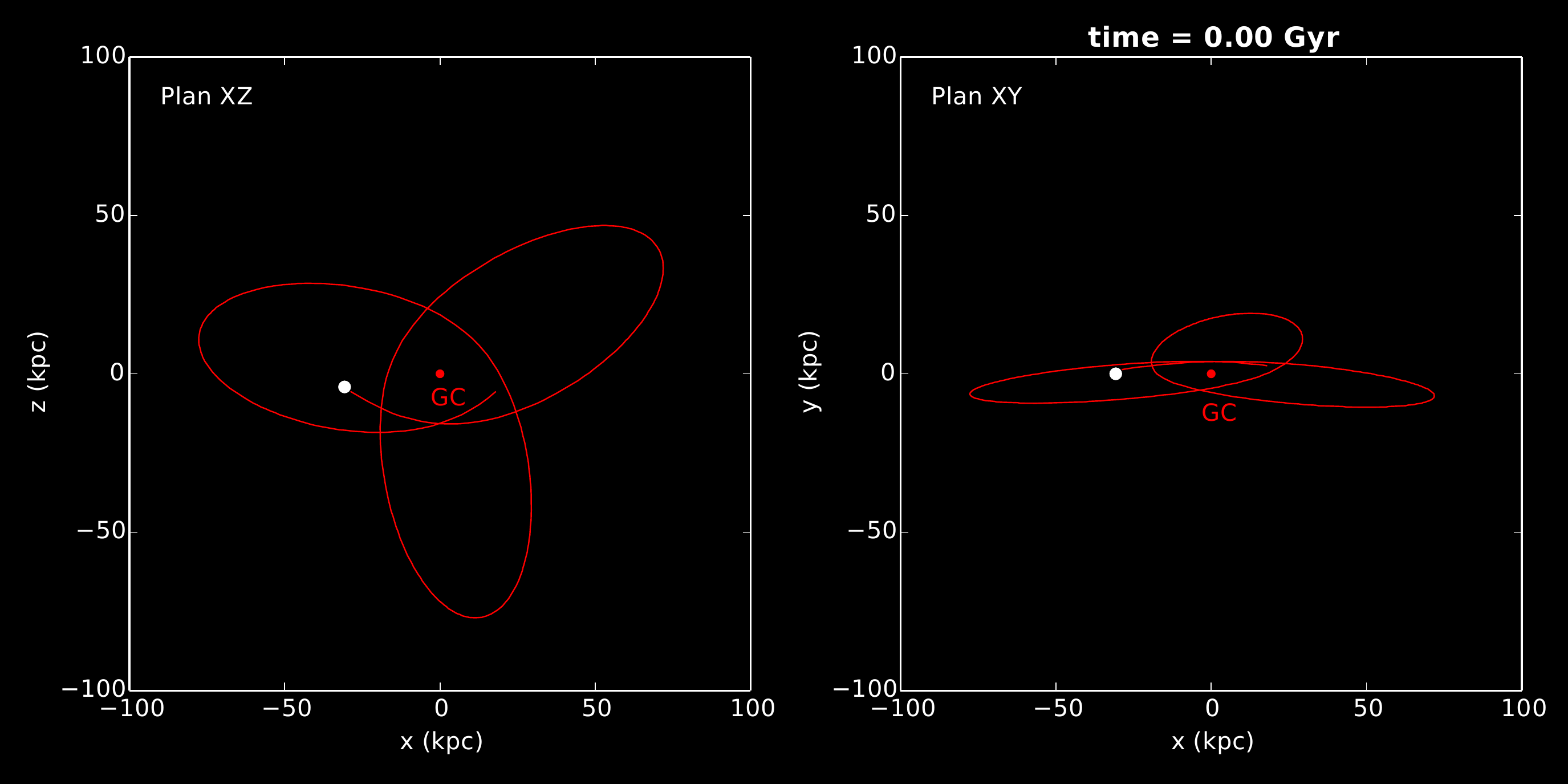}
  \includegraphics[angle=0, viewport= 10 15 1140 590, clip, width=17cm]{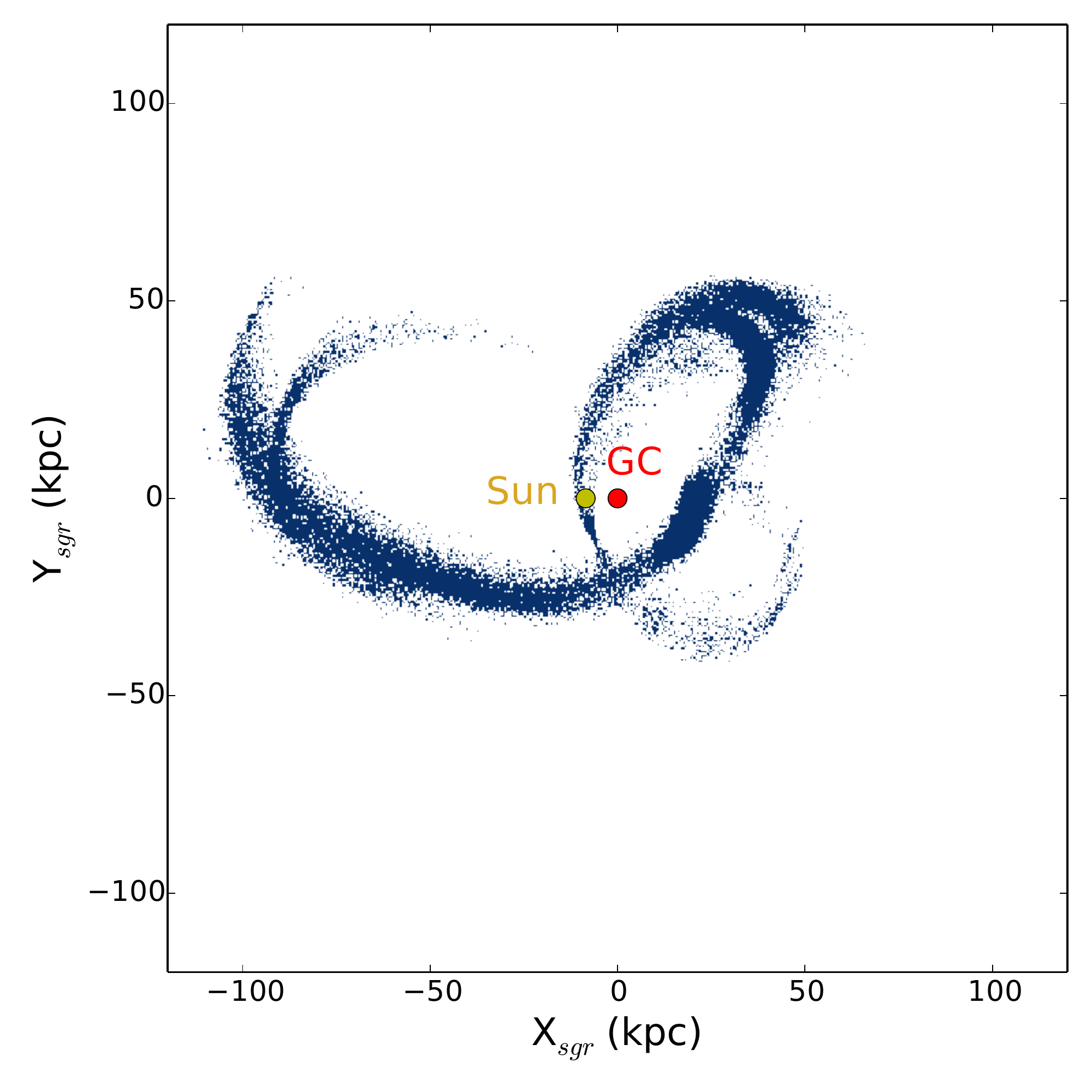}
  \caption{The two upper panels display the movie of the disruption of the Sgr dwarf in our fiducial MOND simulation M1 (The movie can be seen at this \href{http://astro.unistra.fr/fileadmin/upload/DUN/observatoire/Images/GFThomas_MONDSgrstream_movie.mp4}{link}). The top-left panel is a view in the equatorial plane, and the top-right panel is a view from the north Galactic pole. The position of the Sun {\it at the present time} is in yellow. The coordinates are Galactocentric. The lower panel shows the projection at the present time of our M1 model in the plane of the Sgr stream as defined in the left panel of Figure 10 from \citet{belokurov_2014}.}
\label{movie_simple}
\end{figure}

\begin{figure*}
\centering
  \includegraphics[angle=0, viewport= 10 15 1140 570, clip, width=17cm]{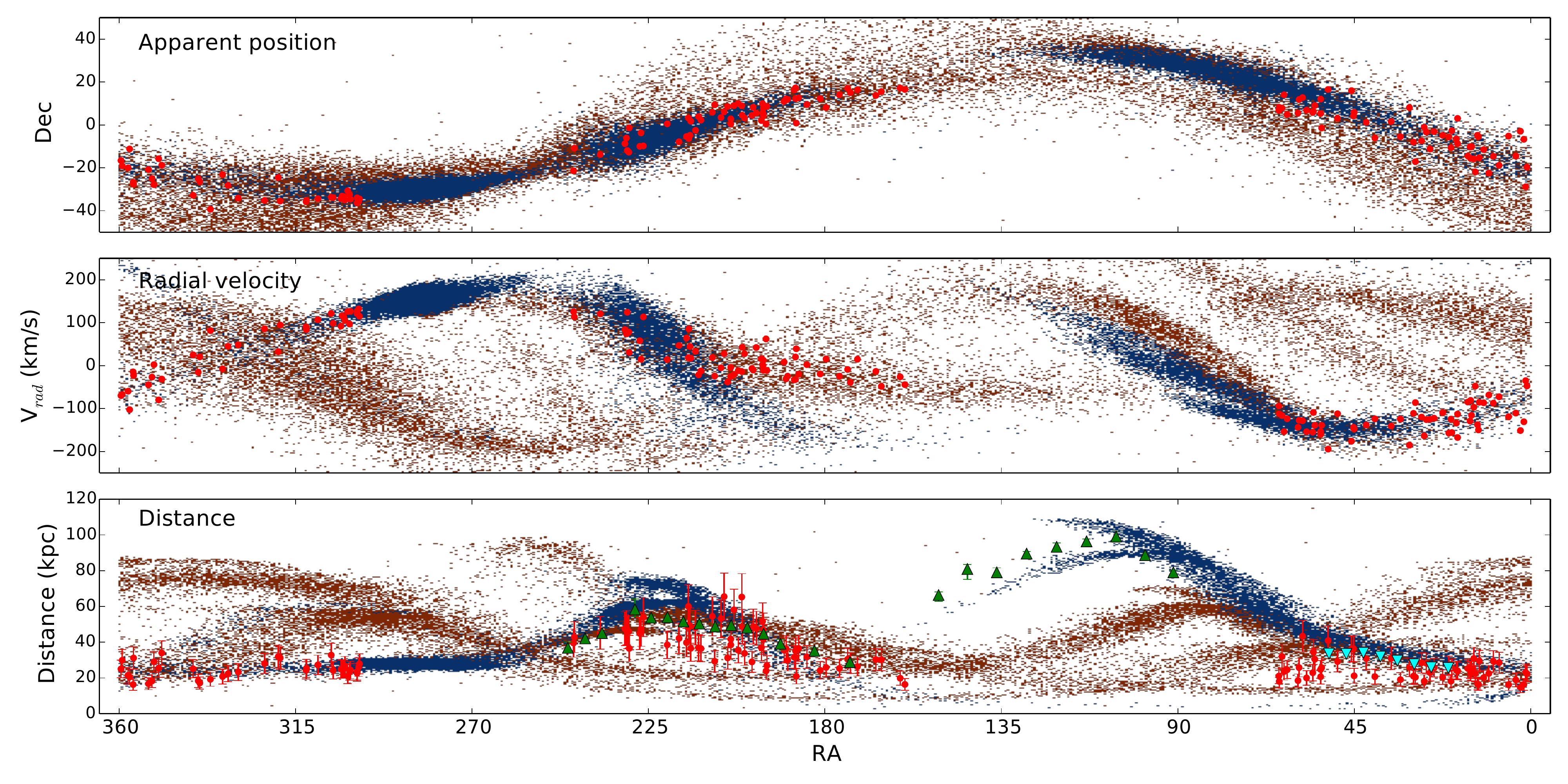}
  \caption{ As Fig. \ref{proj_spherique_dm}, but showing the MOND disk model simulation M1 in blue and the simulation of \citet{law_2010} in orange.}
\label{proj_dsph}
\end{figure*}

Let us insist here on the very little wiggle room we had to choose these initial conditions. The MW potential is entirely determined by its baryonic distribution and so are the initial positions and velocities. The Sgr initial model is only made of stars and should fall on the stellar-mass size relation for similar isolated dwarfs. It is then truly remarkable that the morphology of the remnant after 4~Gyr of disruption is in perfect agreement with that observed by \citet{majewski_2003} with the 2MASS survey. In Fig. \ref{profil_remnant} we compare the density of stars along the minor axis in our M1 simulation (in blue) and in the observed M-giants (in red). The projected minor axis is least affected by tidal effects, and is thus the best direction to evaluate the morphology of the observed remnant. In the M1 simulation, our remnant has a final mass $M_{\rm final} =5.1 \times 10^{7}$, a position angle of $104 \degr$, a half-light radius along the minor axis of $r_h = 0.64$ kpc and a central velocity dispersion of $\sigma_c = 11$ ${\rm km} \, {\rm s}^{-1}$ (see Table~\ref{param_progenitor}), which is in excellent agreement with observations, (see Table~\ref{param_Sgr}) assuming a mass-to-light ratio $\gamma_* = 2.1$ in the \textit{V}-band. In Fig. \ref{surface} we also show the surface brightness of the remnant for the M1 simulation on the upper panel. This morphology can be directly compared to Figure~4 of \citet{majewski_2003}.

In Fig.~\ref{movie_simple}, we show a movie of the M1 simulation, and also provide a projection of the final shape of the stream at present time in the Sgr orbital plane defined by \citet{belokurov_2014} (note that the x-axis is points in the opposite direction to that in their paper). We note that the modelled stream extends to slightly larger distances than 100 kpc, but not much beyond that. This could be different in a simulation where the MW model itself would be live and responding to the gravitational pull of the dwarf. We also display the resulting projected positions on the sky, radial velocities and distances of our M1 simulation in Fig.~\ref{proj_dsph}, together with the Newtonian simulation of \citet{law_2010}. The positions and distances of the M1 stream are in reasonable agreement with the observations. But again, the radial velocities do not match well in the leading arm, exactly as in simulation N1 and in all Newtonian models with spherical, oblate, or prolate halos \citep[e.g.][]{law_2005, dierickx_2017}. While this could be due to the influence of other satellites such as the LMC, we consider hereafter another possible solution based on the influence of the hot gas corona around the Galaxy. We note that the M1 stream is actually very similar to that produced in the N1 simulation (see Fig. \ref{proj_spherique_dm}, the small differences are of the same scale as the resolution of the grid of the RAMSES code), except that it is more self-consistent as the progenitor consists only of stars, and is more constrained in the sense that the progenitor had to obey observed scaling relations. This similitude with the N1 stream means that the external field effect does not have an important effect on the morphology of the stream in MOND, and is dominated by the effective gravitational potential of the MW at the distances probed by the orbit of Sgr. 

\subsection{The Sgr stream in the hot corona model} \label{result_HG}

Here we present our second MOND simulation, denoted M2, in which a massive diffuse hot gas (HG) corona is included. The presence of such a hot diffuse gaseous corona around the MW at a temperature of $\sim 10^6$ K has been proposed for many years as a significant reservoir of baryonic matter that can be traced by the O VII and O VIII emission and absorption lines \citep{paerels_2003} in the soft X-ray band. Recent measurements obtained with the XMM-Newton and Suzaku X-ray telescopes have estimated a mass for this hot gaseous spherical component between $(0.5 \, $-$ \, 1) \times 10^{11}$ M$_\odot$, and could even be as massive as $1.5 \times 10^{11}$ M$_\odot$ \citep[but see ][for lower estimates]{gatto_2013,salem_2015}, while it should extend up to at least 100 kpc \citep{gupta_2012,fang_2013}. Such a corona may actually be the remnant of the formation of the vast polar structure of satellite galaxies (VPOS) if it consists of dwarf galaxies and star clusters that formed within a large gas-rich tidal arm which was pulled out about 10 Gyr ago of either the young MW or the passing other galaxy which may have been Andromeda \citep{pawlowski_2011,zhao_2013,hammer_2013}.  The star-formation efficiency is at most a few per cent on the scales of molecular clouds and within a tidal arm of extend of 100 kpc or more it is likely to have been significantly smaller, probably less than 0.1 per cent. The stellar mass in the VPOS comprises about $10^{9}\,M_\odot$ in total, such that this very rough estimate implies about $10^{11}\,M_\odot$ to have been the mass of the tidal arm(s) within which formed the present-day VPOS constituents. Today the gas is likely to be oriented in a thick oblate structure aligned with the VPOS, partially being derived also from ram-pressure stripping and gas blown out from the young tidal dwarf galaxies which were the precursors of the present-day satellite galaxies. Note that the LMC alone could be the progenitor of this corona, as small disk galaxies are very gas rich. The corona is likely to be hot as its low density implies a long cooling time and possibly heating agents such as the motions of the satellite galaxies and intergalactic radiation may continue to heat this oblate corona. The existence of such an ancient  structure which would be the remnant of the tidal arm(s) is rather speculative though,  and theoretical research is needed to provide constraints on the stability and existence of such a structure, if it can be created from a tidal arm at all. It is therefore rather interesting that observational evidence has appeared which suggests the existence of a hot corona, and our M2 model here shows that such a structure significantly improves the reproduction of the Sgr orbit and its tidal arms.

\begin{figure}
\centering
  \includegraphics[angle=0, viewport= 10 15 565 276, clip, width=8cm]{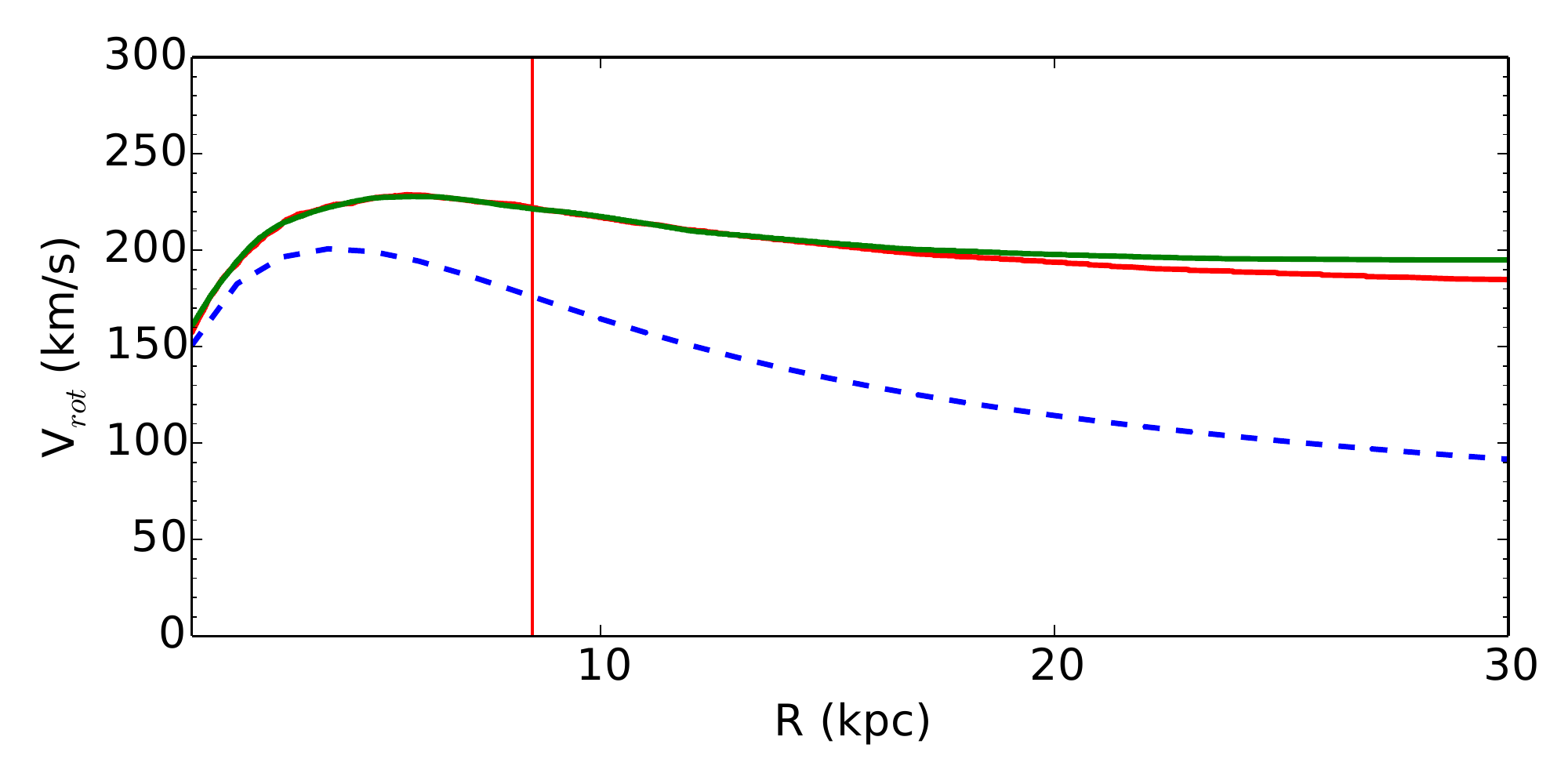}
    \includegraphics[angle=0, viewport= 10 15 565 530, clip, width=8cm]{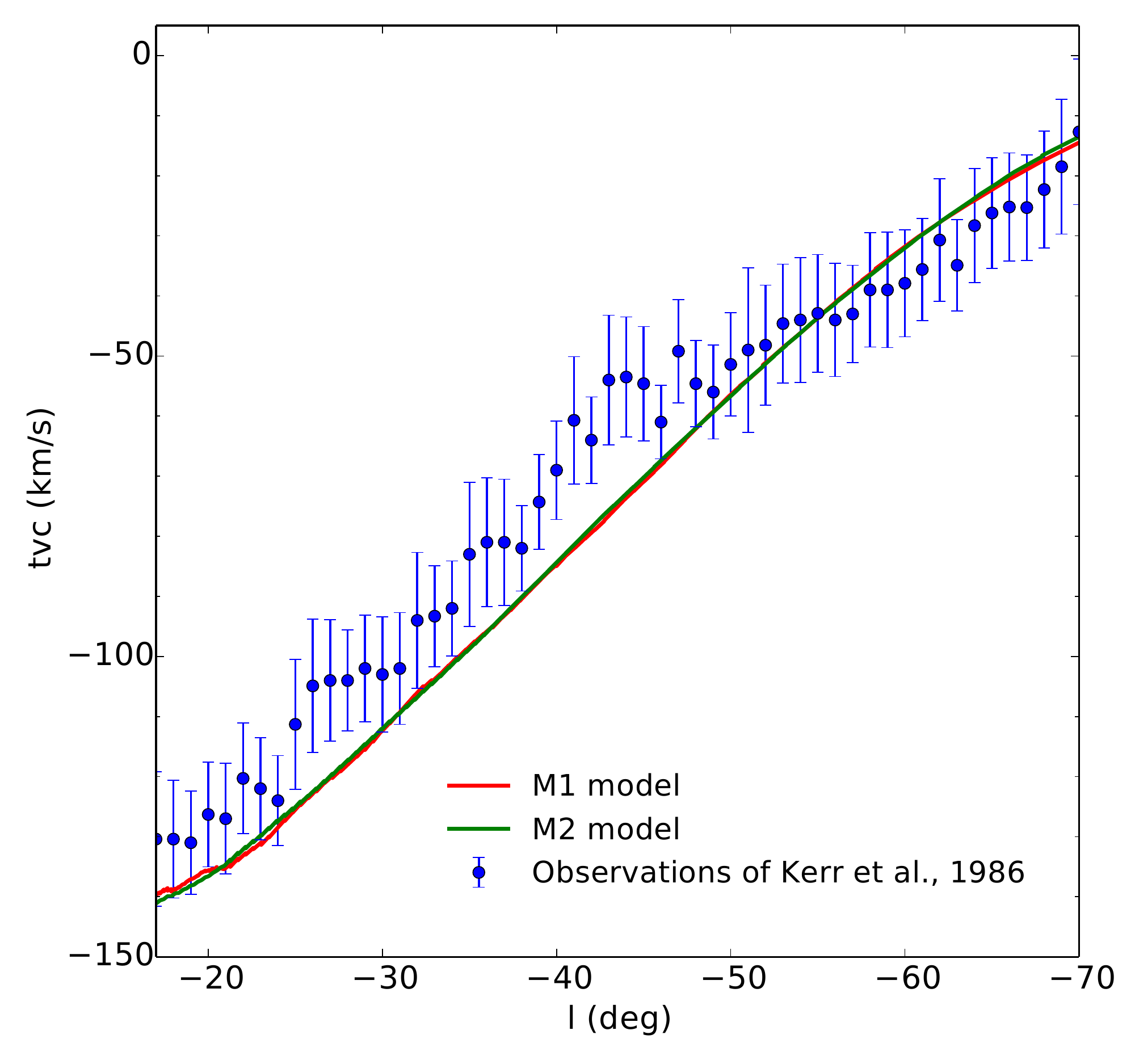}
  \caption{ Top: rotation curves of our models of the MW in MOND along the GC -- Sun axis. The M1 model rotation curve is in red, and the M2 model in green. The difference is clearly very mild. The dashed blue line represents the Newtonian rotation curve without DM and without a hot corona. We fixed the Sun at 8.5 kpc from the GC illustrated by the vertical red line on this figure. Bottom: reproduction of figure 2 of \citet{famaey_2005} for the terminal velocity curve of the M1 and M2 models in the inner Galaxy (fourth quadrant), together with the data of \citet{kerr_1986}. The model is not a perfect representation of the inner Milky Way, but small changes of the inner baryonic structure within the solar radius do not matter much for the Sgr stream.}
\label{rotation}
\end{figure}

In our simulation we modelled this hot gaseous component with a triaxial cored halo profile: 
\begin{equation}
 \rho_{HG} (m)  = \rho_{0,HG} \, \left( 1+\frac{m}{r_{0,HG}} \right)^{-3} \exp \left( \frac{m^2}{r_{t, HG}^2}\right) \, ,
 \label{density_HG}
\end{equation}
where $r_{0, HG}$ is the core radius, $r_{t, HG}$ the truncation radius, and the oblateness is defined throug the elliptical radius $m$ such that $m^2 = x^2/a^2 + y^2/b^2 + z^2/c^2$ and $a=0.44,\,b=1.0, \,c=1.0$. We restrict the total mass of the flattened corona within 100 kpc to be M$(< \,100 \,kpc) \approx 1.5\times 10^{11} M_\odot$, which parametrizes this profile with a total mass up to $2.6 \times 10^{11}$ M$_\odot$ within the truncation radius, but the mass beyond 100~kpc is not necessary, as cutting off the mass beyond 100 kpc does not influence the stream's formation and kinematics. We also show in  Fig.~\ref{rotation} how little this additional corona affects the MW rotation curve.

Since the total baryonic mass of the MW is higher in this M2 model than in M1, the tidal effects on the Sgr dwarf are stronger, and one needs to increase the mass of the progenitor to keep a realistic remnant after 4~Gyr. This can still be done while remaining consistent with the observed total luminosity of the Sgr structure, by taking an initial mass of $M_{init} =1.4 \times 10^{8} {\rm \, M_\odot}$ , an initial core radius of $r_c = 0.6$ kpc and $W = 5$, corresponding to a central velocity dispersion of $\sigma_c = 25$ km.s$^{-1}$ and a half-light radius $r_h = 0.62$~kpc, which is still in agreement with the mass-size relation of other dwarf galaxies \citep{dabringhausen_2013}. We neglect the dynamical friction due to the gas particles of the HG corona, precisely because these are hot and not very reactive to the perturbation from the dwarf, hence not prone to creating dynamical friction, and also because our Newtonian study has shown that even a massive triaxial DM halo does not affect much a light progenitor in Newtonian dynamics, so we can expect the same for a much less massive corona in MOND.

\begin{figure}
\centering
  \includegraphics[angle=0, viewport= 10 15 565 423, clip, width=8cm]{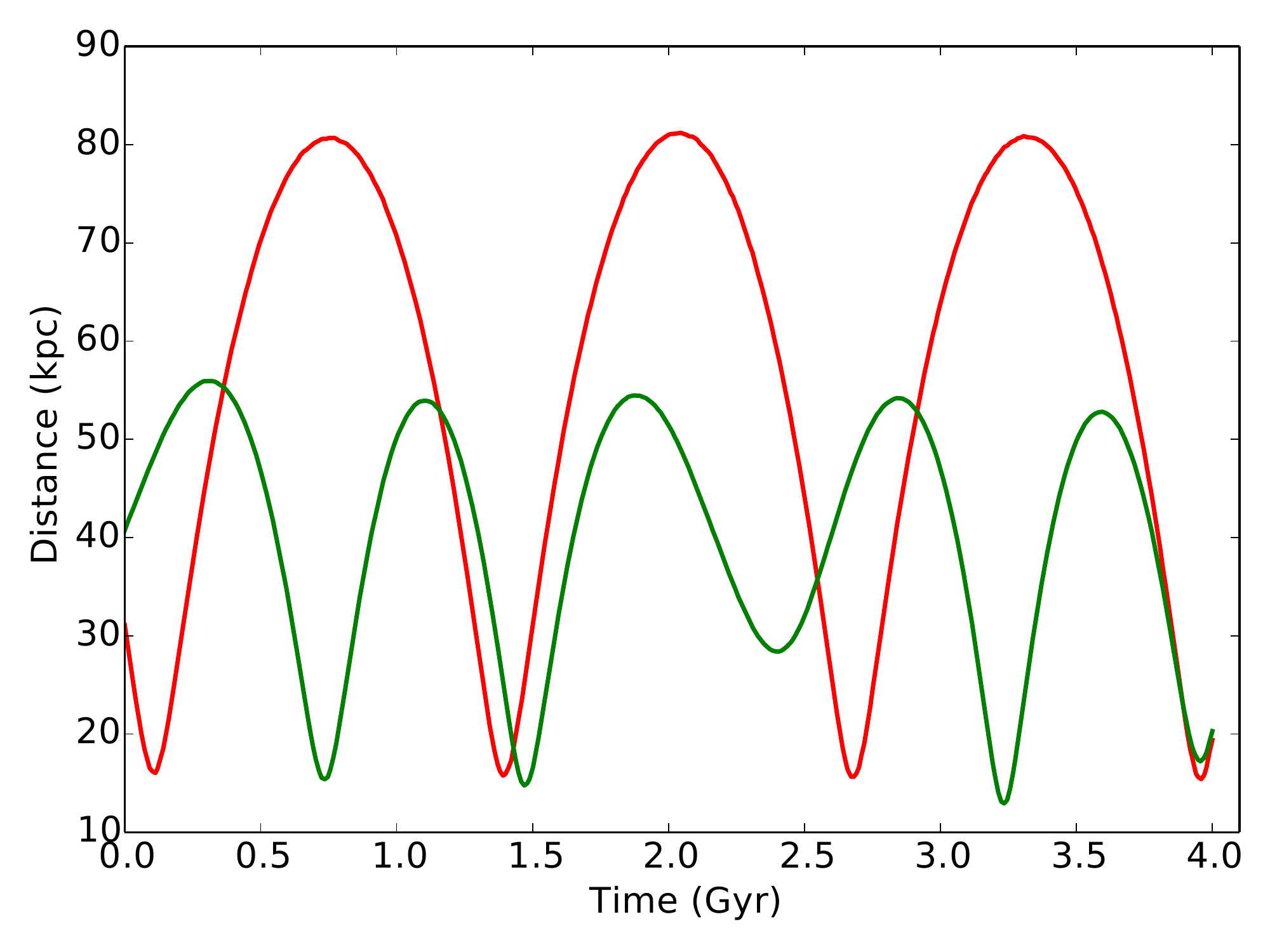}
  \caption{Evolution of the galactocentric distance of the Sgr satellite with the same color code as Fig.~\ref{rotation} (M1=red, M2=green). }\label{orbit_time}
\end{figure}

\begin{figure}
\centering
  \includegraphics[angle=0, viewport= 10 15 565 423, clip, width=8cm]{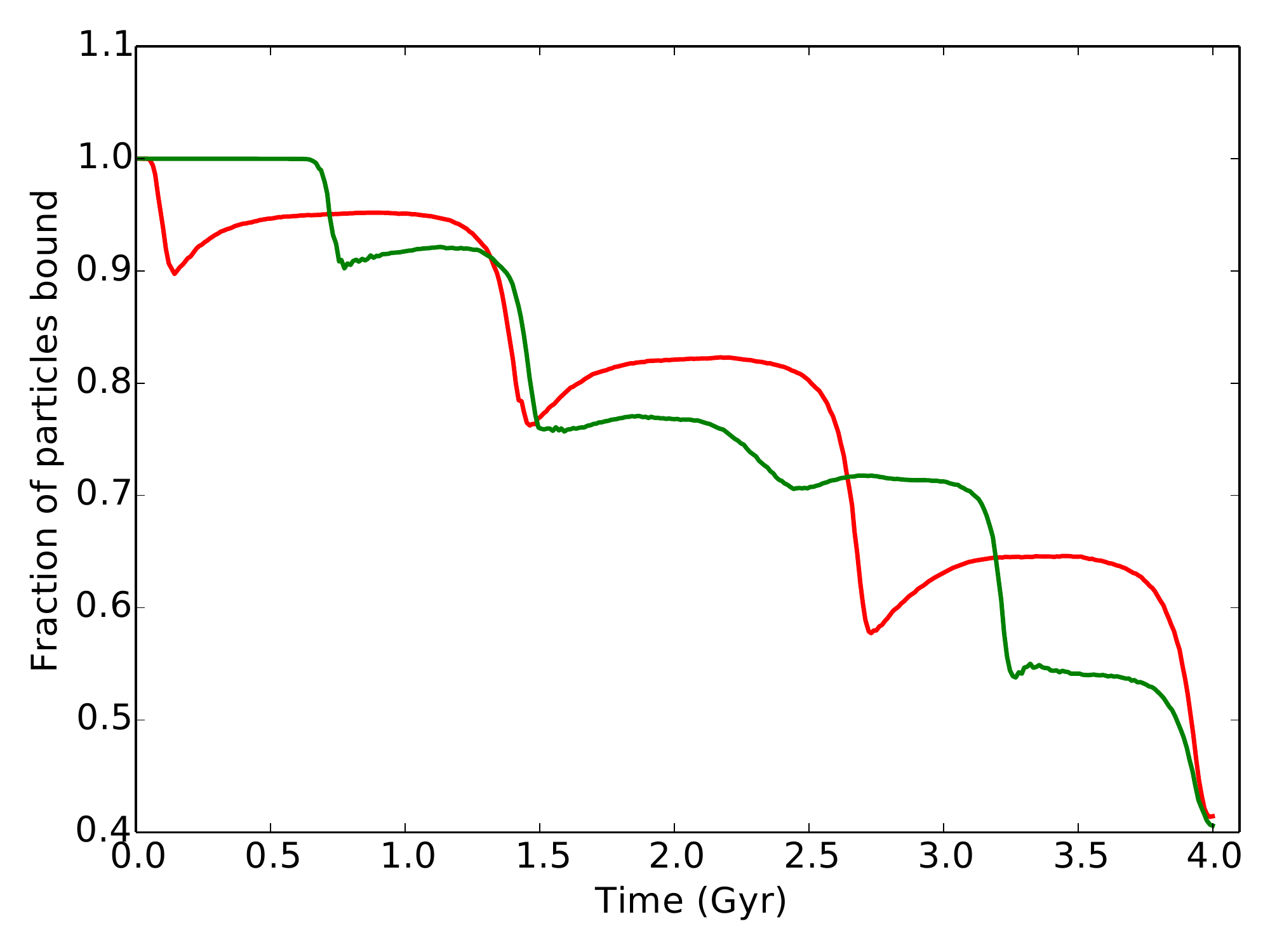}
  \caption{Temporal evolution of the fraction of N-body particles that stay bound to the progenitor (M1=red, M2=green). We checked that the formula of \citet{varghese_2011} for the Jacobi radius was a good approximation in MOND too, and used it to determine if particles are bound. Note how the progenitor in the M1 case significantly increases its mass between $t=2.7$ Gyr and $t=3$ Gyr, through the increase of the Roche radius, allowing it to recover the stars that stay close to it. Note also how the large pericenter of the M2 orbit at $t \approx 2.4$~Gyr limits the mass loss in simulation M2. It is interesting how these effects conspire to give a very realistic remnant in both simulations, as shown in Fig.~\ref{surface}.}
\label{frac_bound}
\end{figure}

After 4 Gyr of disruption, the remnant has a similar morphology to that in the case of the M1 disk model, as can be seen on Fig. \ref{surface} where we show the surface brightness of the remnant for the M1 disk model on the top panel and for the M2 HG model on the bottom panel. The morphology in the two cases is very close to the observed morphology of the Sgr dSph in the 2MASS survey \citep[see e.g. Figure 4 of][]{majewski_2003}. The M2 remnant has a final mass of $M_{\rm final} =5.7 \times 10^{7} {\rm \, M_\odot}$, a position angle of $104 \degr$, a half-light radius along the minor axis of $r_h = 0.66$ kpc and a central velocity dispersion of $\sigma = 11$ ${\rm km} \, {\rm s}^{-1}$ (see Table~\ref{param_progenitor}), that correspond to a stellar mass-to-light ratio of $\gamma_* = 2.4$ to reproduce the observed luminosity of the remnant.

The fact that the remnant is realistic in both the M1 and M2 simulations is interesting, as it results from very different orbital and mass loss histories. The difference between the orbit of the Sgr dSph in the two MONDian models M1 and M2 is shown in Fig \ref{orbit_time} where the orbit for the M1 model is in red and the orbit for the M2 model in green. The pericenter in both cases is of about $\sim 15$ kpc but the apocenter is much closer in the case of the HG corona, $55$ kpc instead $80$ kpc for the M1 model. This is a consequence of the more massive baryonic model of the MW, and could be problematic to reproduce tentative pieces of the stream detected at distances of the order of 100 kpc. It is interesting to see that in the case of the M2 simulation, the pericenter after 2.4 Gyr of disruption is as large as $30$ kpc and thus limits the mass-loss of the progenitor for this orbit, as shown in Fig.~\ref{frac_bound}. The bound fraction of particles can also increase with time, which is especially the case for the M1 model. In both the M1 and M2 models, all these effects conspire to give a realistic remnant. Note how the mass of the remnant is predicted to increase again in the future within the M1 simulation.

\begin{figure*}
\centering
  \includegraphics[angle=0, viewport= 10 15 1140 570, clip, width=17cm]{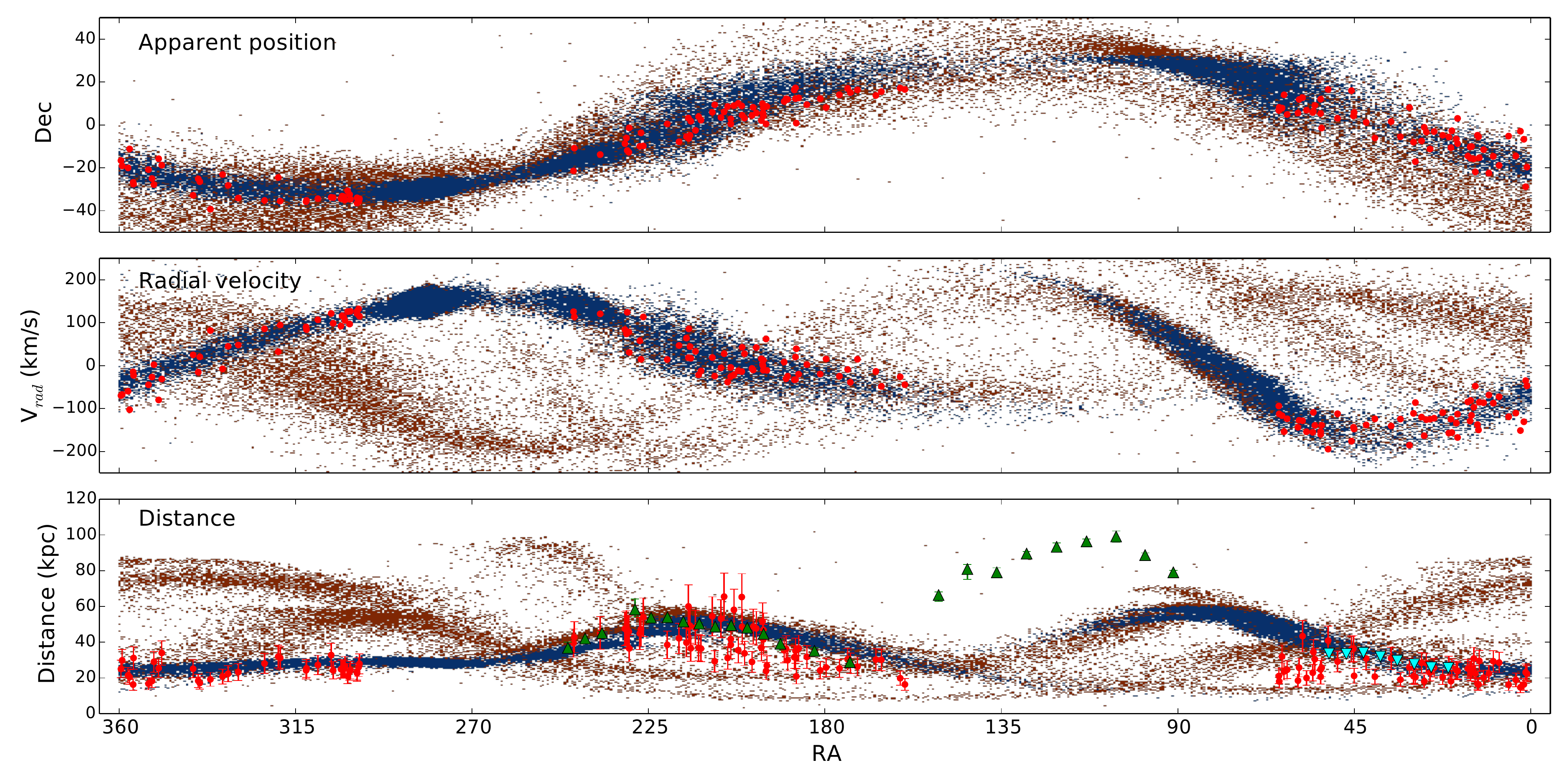}
  \caption{ As Fig.~\ref{proj_dsph}, but for the MOND simulation M2 in blue, and the simulation of \citet{law_2010} in orange. This MOND model (in blue) contains a flattened hot gaseous corona (M$(< \,100 {\rm \,kpc}) \simeq 1.5\times 10^{11} {\rm M_\odot}$) around the MW, aligned with the VPOS.}
\label{proj_HG}
\end{figure*}

Finally, Fig. \ref{proj_HG} represents the observed projected quantities in the same way as Fig. \ref{proj_spherique_dm}. In this M2 model, the simulated radial velocities reproduce very well the observations of the M-giant stars of \citet{majewski_2004} in the leading arm. Contrary to the N2 simulation, the projected dispersion along the stream is also more consistent with the observations, making this model quite superior to that resulting from the LM10 triaxial halo. The much smaller mass of the HG corona, especially in the inner parts, is not likely to destabilize the disk as much as the LM10 DM halo, and could perhaps also be consistent with the Pal~5 stream, which will be the subject of another contribution. However, note that the heliocentric distance extent of the stream in this M2 model is much less than in the M1 model, and it does not fit the BHB stars of \citet{belokurov_2014}, but this is also a shortcoming of the current best simulation in Newtonian gravity to reproduce the leading arm velocities, made by \citet{law_2010}.

 In Milgromian dynamics, the VPOS would likely have formed as a tidal interaction early in the history of our Galaxy, and could have associated with it a much larger component in hot gas than the present-day stellar mass in the VPOS. If this hot gaseous corona shares similar flattening and orientation as the VPOS, our model M2 shows that it would affect the orbit of Sgr and its tidal arms in the right way to make the model consistent with the observed radial velocities of the M-giants near ${\rm RA}=180\deg$. However, we note that recent estimates of the mass of the corona are significantly lower than the mass we assumed here \citep{gatto_2013,salem_2015}. Therefore it is possible that a more realistic model has properties in between M1 and M2.

\section{Discussion and conclusions}

In this paper, we have for the first time rigorously tested the MOND paradigm in a regime where it had never been tested before, namely using stellar streams as a gravitational experiment. We started a series of papers on this topic with an analysis of the most prominent stream of the MW, the Sgr stream. Reproducing such a stream is not a trivial task for a theory such as MOND, as it requires the global shape of the gravitational potential, fully determined by the baryon distribution, to conspire with the tidal effects on the disrupting dwarf galaxy originally (i) sitting on the observed stellar mass-size relation and (ii) reproducing the total luminosity of the stream, to produce both (I) the correct shape and kinematics of the stream, and (II) the correct internal structure and kinematics of the remnant dwarf spheroidal. There was thus very little wiggle room for our simulations to reproduce the observations. 

Starting with a King model of total mass between $1.2 \times 10^{8} M_\odot$ and  $1.4 \times 10^{8} M_\odot$ and half-light radii between 610~pc and 620~pc, in accordance with the observed stellar mass-size relation, our two MOND simulations M1 and M2, with and without a flattened hot gas corona, both gave a very realistic remnant and a quite realistic stream morphology, despite very different orbital and mass-loss histories.

Our M1 model is our fiducial model without hot gas, for which we provided a movie in Fig.~\ref{movie_simple}. This M1 simulation predicts stream debris reaching out to distances slightly larger than 100~kpc, produces a remnant that matches well the observations of the Sgr dSph at the present epoch, and is as well a very good match to the positions of the bright stream arms on the sky. This is quite an achievement for a model with so little freedom. However, the M1 model does not seem to reproduce well the observed radial velocities of M giants in the leading arm, between RA$= 140^\circ$ and $200^\circ$, a well-known problem in Newtonian dynamics for all DM halo models with a spherical, oblate, or prolate shape \citep[e.g.][]{law_2005, dierickx_2017}. 

The similarity of the M1 stream with our Newtonian spherical DM halo model N1 means that the external field effect does not have an important influence on the morphology of the Sgr stream in MOND, and is dominated by the quasi-spherical `phantom DM' halo of the MW at the distances probed by the orbit of Sgr. This will not necessarily be the case for less massive progenitors (such as Palomar~5), which will be the topic of a further paper in this series. 

If the radial velocities in the leading arm are not considered as misidentified stream members, two possible solutions in MOND would be (i) the influence of other satellites such as the LMC \citep[e.g.,][]{vera-ciro_2013, laporte_2016} or (ii) the influence of a flattened hot gas corona aligned with the VPOS. While solution (i) will be the topic of further study, we examined here the plausibility of the solution (ii) in our M2 simulation. In this M2 model with a massive flattened hot gas corona, the simulated radial velocities reproduce very well the observations of the M-giant stars of \citet{majewski_2004} in the leading arm. As opposed to the N2 simulation based on the triaxial DM halo of LM10, the projected dispersion along the stream is also more consistent with the observations in this M2 simulation. Nevertheless the extent of the stream is predicted to be much smaller in this M2 model than in the fiducial M1 case, which could be problematic (checking whether BHB stars at large distances could be fitted in such a model would need to consider a much longer orbit), and the assumed mass of the corona is high compared to some recent estimates. In the future, it should be interesting to test if other configurations of the progenitor or of the flattened hot gaseous corona are also able to reproduce the bifurcation seen in the Northern sky in SDSS and recently extended in the Southern hemisphere by \citet{navarrete_2016}. In the same way, modelling other streams, especially Pal 5, with this hot gaseous corona in MOND should be done in the future, to see if the Pal 5 stream stays coherent in this model, contrary to its structure in the MW potential derived by LM10 \citep{pearson_2015}. It will also be interesting to see if the external field effect plays a more important role for lower mass progenitors such as Pal~5, and could leave a distinctive MOND signature in the stream. All this will be the topic of further papers in this series on using stellar streams as gravitational laboratories.

\section*{Acknowledgments}

The authors would like to thank Stacy McGaugh and Nicolas Martin for very helpful discussions and precious remarks.

  \bibliographystyle{aa}
  \bibliography{./biblio}
\end{document}